\documentclass[3p,12pt]{elsarticle}

\usepackage{amsfonts,amsmath,amssymb,graphicx,graphics}

\usepackage{hyperref}

\usepackage{lineno}

\usepackage{bm}

\usepackage{cases}
\usepackage{array}

\usepackage{booktabs}
\usepackage{multirow}
\usepackage{siunitx}
\usepackage{pifont}
\usepackage{tabularx}
\usepackage{subcaption} 
\usepackage[utf8]{inputenc}
\usepackage{natbib}
\usepackage{mathtools}
\usepackage{physics}
\usepackage{enumitem}
\usepackage[linesnumbered,ruled,noend]{algorithm2e}
\usepackage{xcolor}
\usepackage{newtxtext,newtxmath,amsmath} 

\graphicspath{ {./figures/} }

\def\HiLi{\leavevmode\rlap{\hbox to \hsize{\color{yellow!30}\leaders\hrule height .8\baselineskip depth .5ex\hfill}}}

\journal{}

\begin{document}

\title{A multi-level parallel solver for rarefied gas flows in porous media}

\author[add1]{Minh Tuan Ho\fnref{fn1}}
\author[add1,add2]{Lianhua Zhu\fnref{fn1}}
\author[add1]{Lei Wu}
\author[add1]{Peng Wang}
\author[add2]{Zhaoli Guo}
\author[add3,add4]{Zhi-Hui Li}
\author[add1]{Yonghao Zhang\corref{cor1}}
\ead{yonghao.zhang@strath.ac.uk} 
\cortext[cor1]{Corresponding author}
\fntext[fn1]{Both authors contributed equally.}

\address[add1]{James Weir Fluids Laboratory, Department of Mechanical and Aerospace Engineering, University of Strathclyde, Glasgow G1 1XJ, UK}
\address[add2]{State Key Laboratory of Coal Combustion, School of Energy and Power, Huazhong University of Science and Technology, Wuhan, 430074, China}
\address[add3]{Hypervelocity Aerodynamics Institute, China Aerodynamics Research and Development Center, Mianyang 621000, China}
\address[add4]{National Laboratory for Computational Fluid Dynamics, No.37 Xueyuan Road, Beijing 100191, China}

\begin{abstract}
	A high-performance gas kinetic solver using multi-level parallelization is developed to enable pore-scale simulations of rarefied flows in porous media. The Boltzmann model equation is solved by the discrete velocity method with an iterative scheme. The multi-level MPI/OpenMP parallelization is implemented with the aim to efficiently utilise the computational resources to allow direct simulation of rarefied gas flows in porous media based on digital rock images for the first time. The multi-level parallel approach is analyzed in details confirming its 
	better performance than the commonly-used MPI processing alone for an iterative scheme.
	With high communication efficiency and appropriate load balancing among CPU processes, parallel efficiency of $94\%$ is achieved for $1536$ cores in the 2D simulations, and $81\%$ for $12288$ cores in the 3D simulations. While decomposition in the spatial space does not affect the simulation results, one additional benefit of this approach is that the number of subdomains can be kept minimal to avoid deterioration of the convergence rate of the iteration process. This multi-level parallel approach can be readily extended to solve other Boltzmann model equations.\\


\end{abstract}


%


\begin{keyword}
rarefied gas dynamics, porous media, multi-level parallel, permeability
\end{keyword}

\maketitle

	
\section{Introduction}\label{sec:intro}

Gas transport in ultra-tight porous media has recently received extensive attention due to its important role in extracting unconventional gas resources and developing new technologies such as fuel cells, filters and catalytic converters~\cite{Wang:2014,Harel:2006}. However, understanding of gas transport in ultra-tight porous materials remains a research challenge for theoretical, computational and experimental studies. As the pore space can be as small as a few nanometers, the conventional continuum flow theories such as the Darcy law and the Klinkernberg model become invalid. Meanwhile, the low permeability (at the nano-Darcy scale) is difficult for experimental measurements~\cite{Darabi:2012}. Therefore, direct pore-scale simulation of gas flow in ultra-tight porous material becomes even more important to unravel complex gas transport mechanisms, which can also provide reliable data for upscaling, e.g. the apparent permeability obtained at the representative elementary volume scale can be plugged into macroscopic upscaling equations for predicting gas production~\cite{Lunati:2014}.

In ultra-tight porous media, gas rarefaction effect becomes important to gas transport as the molecular mean free path is comparable to the average pore size. The rarefaction level can be indicated by the Knudsen number $Kn$, defined as the ratio of the molecular mean free path $\lambda$ to the characteristic length~$L$ of the flow, i.e.
\begin{equation}
\begin{aligned}
Kn=\frac{\lambda}{L},\qquad 
\lambda=\frac{\mu (\hat{T}_0)}{\bar{p}}\sqrt{\frac{\pi R\hat{T}_0}{2}},\qquad  
\end{aligned}
\label{eq:Knudsen_number}
\end{equation}
where $R$, $\bar{p}$ and $\mu (\hat{T}_0)$ are the specific gas constant, mean gas pressure and dynamic viscosity of gas at a reference temperature $\hat{T}_0$, respectively. Gas flows can be roughly categorized into four regimes: continuum $(Kn < 0.001)$, slip $(0.001< Kn < 0.1)$, transition $(0.1 <Kn < 10)$ and free-molecular $(Kn > 10)$ flow regimes. The Navier-Stokes equations are applicable for continuum flow regime and their validity might be extended to slip flow regime by introducing velocity-slip boundary conditions at the solid surfaces~\cite{Maxwell:1879,Sharipov:2011}. Further increasing rarefaction level into the transition and free-molecular flow regimes, the linear constitutive relations as assumed in the Navier-Stokes equations are no longer valid~\cite{Lockerby:2008}. Consequently, gas kinetic equations such as the Boltzmann equation or its model equations should be used instead of the Navier-Stokes equation. 

The Boltzmann equation for rarefied gas flows, describes the temporal $\hat{t}$ and spatial $\hat{\mathbf{x}}$ evolution of
the distribution function $\hat{f}(\hat{\mathbf{x}},\hat{\mathbf{v}},\hat{t})$ of gas molecules with the velocity $\hat{\mathbf{v}}$ 
\begin{equation}
\begin{aligned}
\dfrac{\partial \hat{f}}{\partial \hat{t}}+\hat{\mathbf{v}}\cdot\dfrac{\partial \hat{f}}{\partial\hat{\mathbf{x}}} & =I\left(\hat{f},\hat{f}_{*}\right),
\end{aligned}
\label{eq:Boltzmann_equation}
\end{equation}
where the Boltzmann collision integral $I(\hat{f},\hat{f}_{*})$ represents the rate at which the distribution function varies before $(\hat{f})$ and after $(\hat{f}_{*})$ collisions.  Numerical solutions of this nonlinear integro-differential equation for realistic problems 
are tremendously expensive, in terms of computational cost and memory requirement, which is due to the complex five-fold integral $I$ as well as multidimensional phase space (3D spatial domain $\Omega_x$, 3D velocity domain $\Omega_v$, and 1D temporal domain $\Omega_t$ for unsteady flows). The full Boltzmann solvers can be 
classified into
stochastic approach, such as the direct simulation Monte Carlo (DSMC) method, and deterministic approach, such as discrete velocity method (DVM) and fast spectral method (FSM)~\cite{Bird:1994,Pareschi:1996,Kolobov:2007,Kloss:2008,Scanlon:2010,Frezzotti:2011b,Wu:2013}. In order to obtain reasonable approximation 
of the Boltzmann equation,
the full collision integral $I$ is commonly simplified by a relaxation-time collision model, e.g. Bhatnagar-Gross-Krook (BGK)~\cite{Bhatnagar:1954}, ellipsoidal-statistical BGK (ES-BGK)~\cite{Holway:1966} or Shakhov (S)~\cite{Shakhov:1968} model. The Boltzmann model equations preserve the main physical properties, e.g. conservation of mass, momentum and energy, and their accuracy in modeling rarefied gas flows has been assessed. The models have been used for a wide range of applications from high-altitude aerodynamics to microfluidics. 
Among various numerical methods for the Boltzmann model equations, lattice Boltzmann model (LBM)~\cite{Qian:1992} is well developed and widely used for modeling flows in porous media thanks to the ease implementation of boundary condition on complex surfaces~\cite{Blunt:2013,Wang:2016}. 
However, the conventional LBM fails to capture rarefaction effects 
 in early transition flow regime even for a simple Poiseuille flow 
 due to finite number of velocity grid points $N_v$~\cite{Kim:2008, MengJCP11a}. It is demonstrated that high-order LBM is needed to capture the Knudsen paradox phenomena in a straight channel~\cite{Meng:2011b, MengJFM}. Therefore, the accuracy of conventional LBM for porous media flows in the transition and free-molecular regimes is still questionable and other gas kinetic methods 
 are required to simulate rarefied gas flows in these regimes. As the DSMC method is impractical for often low-speed flows in porous media, the DVM is adopted here.

Even with the most simple collision model, i.e. the BGK equation, multi-dimensional phase space
still prohibits practical simulations from serial computing. For example, a serial calculation of a porous sample of $1000^3$-voxel image may need several terabytes of memory and years of the wall-clock time, even with a modest molecular velocity grid of $N_v=10^3$. In the last decade, rapid advances in massively parallel computing technology have paved the way for gas kinetic solvers for direct simulations of flows in porous media~\cite{Frezzoti:2011,Dimarco:2015,Titarev:2012,Titarev:2014,Titarev:2016,Li:2009,Li:2015,Baranger:2014,Zhu:2017}.
For calculations on a single computing node, kinetic solvers are usually parallelized using Open Multi-Processing (OpenMP) or Graphical Processing Unit (GPU). The OpenMP of shared memory model is accomplished by multi-threaded parallelism with the advantages of simplicity, portability and extensibility. However, scalability of OpenMP is limited by the number of cores on a single computing node. For example, speedup of $46$ times is reported for 3D Sod problem executed on a decent 64-core 
machine~\cite{Dimarco:2015}. 
For applications that seek floating point performance, GPU is much faster than the traditional CPU (Central Processing Unit). 
The GPU-based BGK solver may provide impressive acceleration (about 600 and 340 times for the 1D shock wave and 2D driven cavity tests, respectively) thanks to hundreds or thousands of "light weight" stream processors of GPU
~\cite{Frezzoti:2011}. Nonetheless, it is normally restricted to relatively small domain mainly due to limited on-board global memory, e.g. a total grid points of up to $N_x \cdot N_v=25.6\times10^{3}\cdot 8\times10^{3}$ is reported in the above study.

For large scale calculations, the computing nodes are interconnected by high-speed networks and the Message Passing Interface (MPI) library is usually utilized for communication between subdomains, which are decomposed in either the velocity domain $\Omega_v$ or the spatial domain $\Omega_x$. 
The $\Omega_v$-decomposition is simple to implement and easy to achieve load balancing, thus it is usually favoured. For instance, in calculation of 3D hypersonic flows around space vehicles, in which high resolution and a wide range of velocity space are desired, the $\Omega_v$-decomposition strategy enables high efficiency of $88 \%$ upto ${20000}$-core computing (in comparison with the 500-core one) with the total grid points of $191\times10^{3}\cdot 729\times10^{3}$~\cite{Li:2015}. 
Nevertheless, large amount of communications in calculating macroscopic parameters (i.e. the moments of distribution function $\hat{f}$ over the velocity domain $\Omega_v$) at every $N_x$ spatial grid points reduce  the performance significantly, especially for a large $N_x$. 
Hence $\Omega_x$-decomposition is preferred (or inevitable) when the spatial domain is large compared to the velocity domain
. For example, the simulations of 3D pressure-driven flow in a pipe, which uses nearly $350\times10^{3}\cdot 4.1\times10^{3}$ grid points, have parallel efficiencies of $96 \%$ for $\Omega_x$-decomposition and $70 \%$ for $\Omega_v$-decomposition using 512 cores (in comparison with the 64 cores)~\cite{Titarev:2014}. 
Regardless of $\Omega_x$-decomposition or $\Omega_v$-decomposition, pure MPI parallelization has limited scalability for a fixed number of grid points $N_x \cdot N_v$. The subdomain handled by each MPI core becomes smaller whereas the total amount of MPI communications increase with the number of core $N_c$, leading to 
load unbalancing.
One can observe significant drop of efficiency when the number of cores doubles to 1024 in the above example, i.e. from $96 \%$ to $64 \%$ and from $70 \%$ to $15 \%$ respectively. Performance comparison between the two strategies of domain decomposition can be found in Ref.~\cite{Zhu:2017}. 

The limitation on MPI scalability can be mitigated by introducing the second level parallelization. More precisely, two-level MPI/OpenMP parallel approach, in which the top-level MPI proccess is responsible for each $\Omega_x$-subdomain, and a group of OpenMP threads at the bottom-level accelerate computing within the $\Omega_x$-subdomain as well as communicating with other $\Omega_x$-subdomains. This two level approach will help to achieve load balancing and reduce memory consumption for duplicated data~\cite{Rabenseifner:2009}. 
To the best of our knowledge, this multi-level parallel approach for a gas kinetic solver was first proposed by Baranger et al.~\cite{Baranger:2014}. 
However, the tested largest domain is relatively small with the total grid points of $50\times10^{3}\cdot 3\times10^{3}$ and the parallel performance was not analyzed in that study. A typical modern micro-CT scanner can provide images with $2000^3$ voxels and advanced synchrotron imaging may allow much larger images to be reconstructed~\cite{Blunt:2013}. As a result, the two-level parallel MPI/OpenMP approach with excellent scalability is required for the kinetic solver to be able to directly simulate gas flows in the porous media with flow passages provided by high-resolution images. 

Apart from parallel scalability, convergence rate to steady state also contributes to the total simulation time. To solve the BGK equation, the most numerical schemes are designed for unsteady flows~\cite{Yang:1995,Frezzoti:2011,Dimarco:2015,Li:2009,Zhu:2017,Mieussens:2000,Xu:2010,Filbet:2010,Guo:2013}. A few BGK solvers recently employ implicit time-marching scheme to accelerate steady state solutions with a large Courant–Friedrichs–Lewy (CFL) number~\cite{Titarev:2012,Baranger:2014,ZhuY:2016}. It is demonstrated in Refs.~\cite{Sharov:2000, Titarev:2014} that domain decomposition in the spatial space $\Omega_x$ deteriorates, to some extent, the convergence rate of implicit time-marching schemes. Taking into account a smaller number of subdomains for the same number of cores, multi-level parallel approach may mitigate the deterioration in convergence rate for pure MPI $\Omega_x$-decomposition. 

The present work is to develop a multi-level parallel BGK kinetic solver
with enhanced scalability for steady rarefied gas flows in porous media. The solver is designed to efficiently solve flows in porous media with complex structures described by binary 2D slide-images (composed of pixels) or 3D volume-images (composed of voxels). Parallel scalability of two-level MPI/OpenMP approach is analyzed and compared with that of the pure MPI approach.  
   
\section{Governing equation and numerical method}\label{sec:governing_BC}
\subsection{Governing equation and its boundary conditions}
The BGK model equation can be employed to describe low-speed gas flows in ultra-tight porous media, where the flow resistance is high. The distribution function is linearized in the standard manner as $\hat{f}=\hat{f_{eq}}(1+h)$~\cite{Sharipov:1998}, and the Maxwellian distribution function is
\begin{equation}
\begin{aligned} \hat{f}_{eq}=\dfrac{\hat{n}_{eq}}{(2\pi R\hat{T_0})^{3/2}}\exp(-\dfrac{\lvert \hat{\mathbf{v}}\rvert^2}{2R\hat{T_0}}). \end{aligned}
\label{eq:Equilibrium_distribution_function}
\end{equation}
The global equilibrium number density $\hat{n}_{eq}$ is related to the mean gas pressure by the ideal gas law $\hat{n}_{eq}=\bar{p}/mR\hat{T}_0$, in which $m$ is the molecular mass. The (dimensionless) perturbated distribution function $h(\mathbf{x},\mathbf{v})$ is governed by the linearized BGK equation
\begin{equation}
\begin{aligned}
\mathbf{v}\cdot\dfrac{\partial h}{\partial\mathbf{x}}
=\dfrac{\sqrt{\pi}}{2Kn}\left[\varrho+2\mathbf{u}\cdot\mathbf{v}+\tau \left(\lvert \mathbf{v}\rvert^2-\dfrac{3}{2}\right)-h\right],\end{aligned}
\label{eq:linearised_Boltzmann_equation}
\end{equation} 
where time-dependent derivative in Eq.~\eqref{eq:Boltzmann_equation} is omitted as only steady-state solution is of interest. The following dimensionless quantities (denoted by omitting "hat" in the notations of the corresponding dimensional ones) are used
\begin{equation}
\begin{aligned}
\mathbf{x}=\frac{\hat{\mathbf{x}}}{L},\qquad 
\mathbf{v}=\frac{\hat{\mathbf{v}}}{v_m},\qquad 
f=\frac{\hat{f}}{\hat{n}_{eq}/v_m^3},\qquad 
\end{aligned}
\label{eq:dimensionless_quantities}
\end{equation}
where $v_m=\sqrt{2R\hat{T}_0}$ is the most probable speed. The perturbed number density $\varrho$, velocity $\mathbf{u}$ and temperature $\tau$ are calculated as moments of perturbed distribution function $h$ over the velocity space  
\begin{equation}
\begin{aligned}
\varrho=\int f_{eq}h\mathop{\mathrm{d}\mathbf{v}},\qquad 
\mathbf{u}=\int \mathbf{v}f_{eq}h\mathop{\mathrm{d}\mathbf{v}},\qquad
\tau=\dfrac{2}{3}\int \lvert\mathbf{v}\rvert^2f_{eq}h\mathop{\mathrm{d}\mathbf{v}}-\varrho,
\end{aligned}
\label{eq:perturbed_quantities}
\end{equation}
where the dimensionless equilibrium distribution function is 
\begin{equation}
f_{eq}=\pi^{-3/2}\exp(-\lvert\mathbf{v}\rvert^2).
\end{equation}

Boundary conditions for the BGK equation~\eqref{eq:linearised_Boltzmann_equation} should be specified at the solid surfaces and the outer faces of a porous medium. Gas-surface interaction is modeled by the Maxwell diffuse-specular reflection    
\begin{equation}
\begin{aligned}
h\left(\mathbf{v} \mid \mathbf{v} \cdot \mathbf{n} > 0 \right)=\alpha \varrho_s\left( \mathbf{n}\right)
+ \left(1-\alpha\right)h\left( \mathbf{v}-2\mathbf{n} (\mathbf{v} \cdot \mathbf{n}\right)),
\end{aligned}
\label{eq:diffusive-specular_BC}
\end{equation}
where $\mathbf{n},\varrho_s,\alpha$ are the outer normal unit vector of the solid surface, the perturbed gas number density on the solid surface and the tangential momentum accommodation coefficient (TMAC), respectively. Value of TMAC represents the diffuse portion of the reflected molecules, i.e fully diffuse or fully specular reflections correspond to $\alpha=1$ or $\alpha=0$. This study uses the diffuse boundary condition $\alpha=1$ at the solid surfaces. The perturbed gas number density on the solid surface is computed from the non-penetration condition, i.e. zero-mass flux through the solid surface
\begin{equation}
\begin{aligned}
\varrho_s\left( \mathbf{n}\right)=-\dfrac{\int_{\mathbf{v} \cdot \mathbf{n} < 0}\mathbf{v}\cdot \mathbf{n}   \exp(-\lvert \mathbf{v} \rvert ^2)h \mathop{\mathrm{d}\mathbf{v}}}{\int_{\mathbf{v} \cdot \mathbf{n} > 0}\mathbf{v}\cdot \mathbf{n}   \exp(-\lvert \mathbf{v} \rvert ^2) \mathop{\mathrm{d}\mathbf{v}}}.
\end{aligned}
\label{eq:wall_density}
\end{equation}

A porous medium can be constructed by the periodic replica of the representative elementary volume in the $x_1$, $x_2$, and $x_3$ directions, respectively. It is convenient to take the size of computational domain i.e. the representative elementary volume in the $\hat{x}_1$ direction as characteristic flow length $L=\hat{x}_{1}^{max}-\hat{x}_{1}^{min}$
in the definition of Knudsen number Eq.~\eqref{eq:Knudsen_number}. At the inlet ($x_1=x_1^{min}$) and outlet ($x_1=x_1^{max}$), periodic condition~\cite{Sharipov:2012c} representing the pressure gradient is applied for molecules entering the computational domain assuming that the pressure gradient only exists in the $x_1$-direction, i.e. 
\begin{equation}
\begin{aligned}
h\left(x_1^{min},x_2,x_3,v_1,v_2,v_3 \right) &=& (x_1^{max}-x_1^{min}) + h\left(x_1^{max},x_2,x_3,v_1,v_2,v_3 \right),\quad
\text{when} 
\enskip v_1 > 0, \\
h\left(x_1^{max},x_2,x_3,v_1,v_2,v_3 \right) &=& (x_1^{min}-x_1^{max}) + h\left(x_1^{min},x_2,x_3,v_1,v_2,v_3 \right),\quad
\text{when} 
\enskip v_1 < 0.
\end{aligned}
\label{eq:periodic_BC}
\end{equation} 
At the lateral faces of the porous medium, symmetric boundary conditions are implemented by the specular reflection, i.e.    
\begin{equation}
\begin{aligned}
h\left( x_1,x_2^{min},x_3,v_1,v_2,v_3 \right)  &=& h\left( x_1,x_2^{min},x_3,v_1,-v_2,v_3 \right),             \quad \text{when}  \enskip   v_2 >0,  \\ 
h\left( x_1,x_2^{max},x_3,v_1,v_2,v_3 \right)  &=& h\left( x_1,x_2^{max},x_3,v_1,-v_2,v_3 \right),             \quad \text{when}  \enskip   v_2 < 0, \\
h\left( x_1,x_2,x_3^{min},v_1,v_2,v_3 \right)  &=& h\left( x_1,x_2,x_3^{min},v_1,v_2,-v_3 \right),             \quad \text{when}  \enskip  v_3 > 0,  \\ 
h\left( x_1,x_2,x_3^{max},v_1,v_2,v_3 \right)  &=& h\left( x_1,x_2,x_3^{max},v_1,v_2,-v_3 \right),             \quad \text{when}  \enskip   v_3 < 0.      
\end{aligned}
\label{eq:symmetric_BC}
\end{equation}

The dimensionless apparent permeability $k$, which is normalized by $L^2$, is calculated as 
\begin{equation}\label{eq:k_Gp}
k  = \frac{2}{\sqrt{\pi}}KnG_p, 
\end{equation}
where $G_p$ is the dimensionless mass flow rate per unit area normalized by $\bar{p}/v_m$
\begin{equation}
G_p =4 \int_{0}^{1/2} \int_{0}^{1/2} u_1\left(x_2,x_3\right) \mathop{\mathrm{d}x_2} \mathop{\mathrm{d}x_3}.      
\end{equation}

In order to appropriately compare the pore-scale rarefaction effects in different porous samples of various sizes, the effective Knudsen number $Kn^*$ is defined as
\begin{equation}
Kn^*=\frac{\lambda}{L^*}=\frac{Kn}{L^*/L},
\end{equation}
where the average pore size $L^*$ is defined as
\begin{equation}
\begin{aligned}
L^*/L=\sqrt{\frac{12k_\infty}{\epsilon}} \ \ \text{for 2D},\quad   
L^*/L=\sqrt{\frac{8k_\infty}{\pi\epsilon}} \ \ \text{for 3D}.
\end{aligned}
\label{eq:effective_Kn}
\end{equation}
In this study, permeability obtained with the smallest examined $Kn$, which ensures that flow is in the continuum regime, is considered as intrinsic permeability $k_\infty$ to determine $L^*$. The porosity $\epsilon$ of a porous model in Eq.~\eqref{eq:effective_Kn} is determined by the ratio of the number of fluid points to the total number of points, i.e. the percentage of voids in the digital images.  


\subsection{Discrete velocity method}

DVM is one of the most common deterministic approaches to solve the Boltzmann equation and its simplified models~\cite{Broadwell:1964a,Yang:1995}, which projects the continuous molecular velocity space $\mathbf{v}$ into a set of fixed $N_v$-discrete velocities $\mathbf{v}^{(k)}$ $(k=1,2,..,N_v)$. Consequently, the BGK  equation~\eqref{eq:linearised_Boltzmann_equation} is replaced by a system of $N_v$-independent equations. Generally speaking, choice of velocity grid, i.e. the number $N_v$ and value of discrete velocity points $\mathbf{v}^{(k)}$, is problem dependent. The range of $\mathbf{v}^{(k)}$ must cover the scope of bulk velocity $\mathbf{u}$ and temperature $T$ in the whole flow domain, e.g. a broad variety of $\mathbf{u}$ or $T$ requires a wide range of $\mathbf{v}^{(k)}$. The number $N_v$ associates with accuracy order of numerical quadrature adopted for evaluating macroscopic parameters using Eq.~\eqref{eq:perturbed_quantities}. Highly rarefied (large $Kn$) flows usually demand a fine velocity grid to capture the discontinuity in the distribution function. The values of discrete velocity points are determined by the abscissas of the adopted quadrature rule. Detailed discussion on velocity grid can be found in Ref.~\cite{Baranger:2014}. In this study, the Cartesian velocity grid generated by half-range Gauss-Hermit quadrature~\cite{Gautschi:1994} is employed.

\subsection{Finite difference approximation of advection terms}
As reconstructed digital images of porous media samples comprise of voxels, it is straightforward to construct uniform grids, where the fluid and solid points coincident with voids and matrix voxels, respectively.  As a result,  finite difference method (FDM) is convenient to approximate the advection terms on these uniform grids. The spatial derivatives in Eq.~\eqref{eq:linearised_Boltzmann_equation}  are approximated by 
the upwind schemes, e.g. the gradient component of $h$ at the fluid point $\mathbf{x}$ projected in the $x_j$ coordinate axis $(j=1,2,3)$ is evaluated as follows
\begin{equation}
\begin{aligned}
\dfrac{\partial h(\mathbf{x},\mathbf{v})}{\partial x_j}
\approx \mathcal{D}(h(\mathbf{x},\mathbf{v}))_j
= sgn(v_j) [C_{0}h(\mathbf{x},\mathbf{v}) + C_{1}h(\mathbf{x} - sgn(v_j) \Delta x \mathbf{i}_j,\mathbf{v}) + C_{2}h(\mathbf{x} - sgn(v_j) 2\Delta x \mathbf{i}_j,\mathbf{v})],      
\end{aligned}
\label{eq:numerical stencil}
\end{equation}
where $sgn$, $\Delta x$ and $\mathbf{i}_j$ are the sign (signum) function, the constant grid size and the unit vector of the $x_j$ coordinate axis, respectively. The constants $(C_0,C_1,C_2)$ are equal to $(1.5,-2,0.5)/\Delta x$ for the second-order-accurate scheme, while they are $(1,-1,0)/\Delta x$ for the first-order-accurate scheme. The second-order-accurate scheme is used by default while the first-order-accurate scheme is automatically deployed if the upstream grid point in the $v_j$-direction is located on the boundary surfaces. The upwind numerical stencils vary from 3 to 5 points for 2D flows and from 4 to 7 points for 3D problems. For instance, the full 5-point stencil and reduced 4-point stencil for a 2D flow problem are illustrated in Fig.~\ref{fig:halo_sweep_stencil} by the black dash frame around the central fluid point $\mathbf{x}$. To avoid checking the conditional expression \texttt{IF-THEN(-ELSE)} at every iteration, the order of accuracy associated with each spatial grid point is set in pre-processing by storing the values of $(C_0,C_1,C_2)$ in an array. 
Therefore, we obtain a set of discretized equations by applied DVM and FDM on the BGK equation~\eqref{eq:linearised_Boltzmann_equation}  
\begin{equation}
\begin{aligned}
v_j^{(k)}\mathcal{D}(h(\mathbf{x},\mathbf{v}^{(k)}))_j
=\dfrac{\sqrt{\pi}}{2Kn}\left[\varrho+2u_j v^{(k)}_j+\tau \left(\lvert \mathbf{v}^{(k)}\rvert^2-\dfrac{3}{2}\right)-h\right].\end{aligned}
\label{eq:Discretised_linearised_Boltzmann_equations}
\end{equation}

\section{Kinetic solver for rarefied gas flow in porous media}\label{sec:solver_implementation}
\subsection{Algorithmic procedure for the serial solver}
\label{sec:serial_implementation}
\begin{figure*}[thbp]
	\centering
	\includegraphics[width=0.7\textwidth, trim= 20mm 50mm 20mm 60mm]{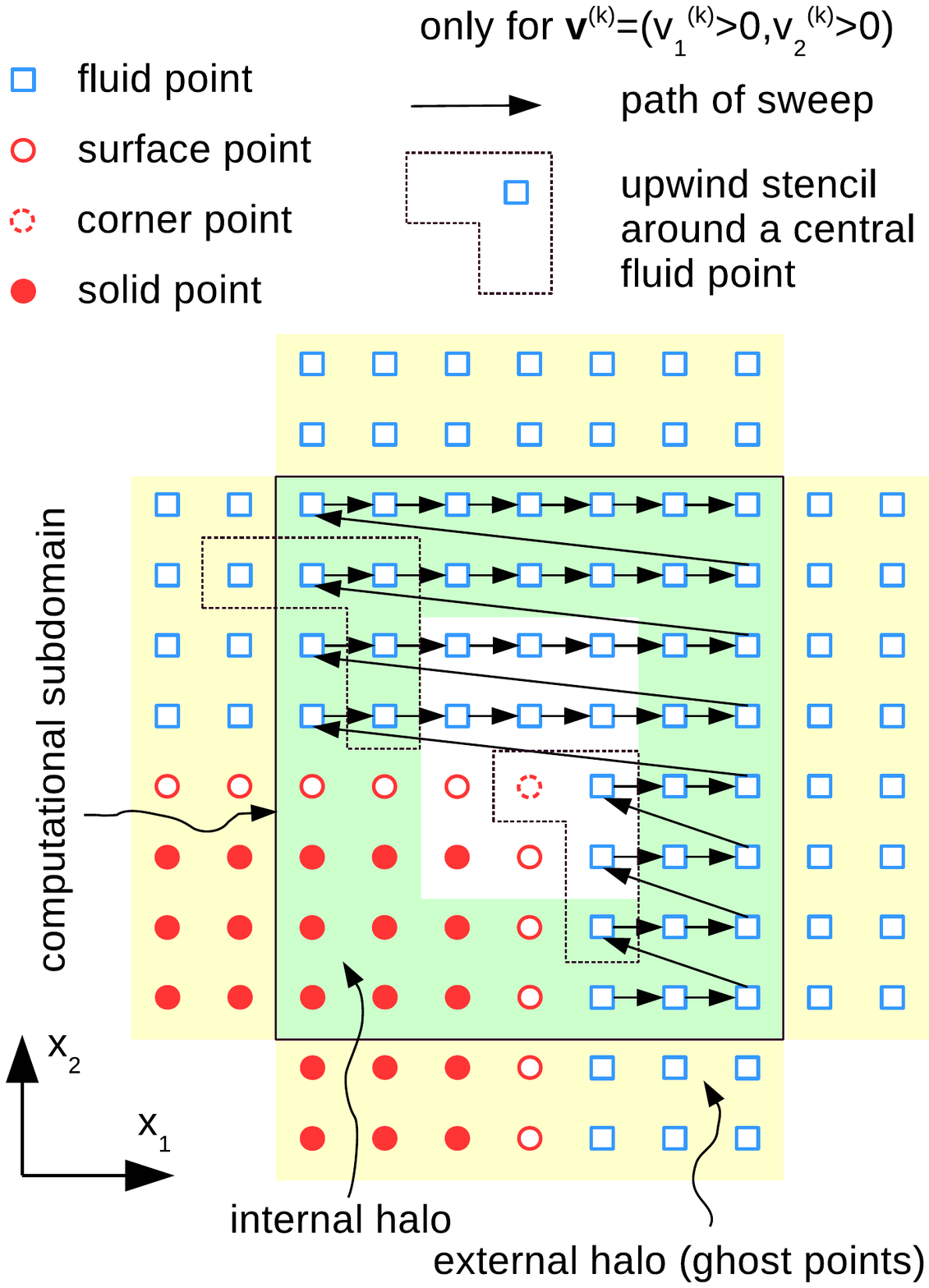}
	\caption{ The subdomain on the top-right quarter of the MPI $2\times 2$ parallelization of the full domain of $14 \times 16$ pixels, in which a square cylinder of $8^2$ pixels is located at the centre. Internal halo/external halo (ghost points), in which data are copied to/duplicated from the neighbouring subdomains (by MPI communication), are in the highlighted regions inside/outside the computational subdomain (bounded by black solid rectangular). Un-highlighted region in the middle holds the data being local to each subdomain, i.e. no MPI communication with others. 
		Forward substitution is performed sequentially by this stencil along a path of sweep, namely raster scan, on all the fluid points inside the computational domain. The upwind stencil (the black dash frame around the central fluid point) and path of sweep of discrete velocity $\mathbf{v}^{(k)}=(v_1^{(k)},v_2^{(k)})$ are subjected to the signs of discrete velocity component, only illustrated for the velocity group of $v_1^{(k)}>0,v_2^{(k)}>0$.}
	\label{fig:halo_sweep_stencil}
\end{figure*}

The set of $N_v$-independent equations, i.e. Eq.~\eqref{eq:Discretised_linearised_Boltzmann_equations}
are solved by an iterative scheme. Three major steps are to be completed successively in each iteration as follows:
\begin{enumerate}[label=(\roman*)]
	\item \label{it:sweep} Distribution function $h$ in the \textit{bulk region} is updated by solving~Eq.~\eqref{eq:Discretised_linearised_Boltzmann_equations}. 
	
	To be more specific, each equation corresponding to a discrete velocity $\mathbf{v}^{(k)}$ can be independently solved by forward substitution along a path of sweep, namely raster scan, on all the fluid points in the spacial domain. The path of sweep is subject to numerical stencil and thus depends on the signs of all the components of the discrete velocity $v_j^{(k)}$.   Figure~\ref{fig:halo_sweep_stencil} demonstrates the path of sweep for the group of discrete velocity $v_j^{(k)}=(v_1^{(k)}>0,v_2^{(k)}>0)$ in a 2D simulation.  This path starts on the  bottom-left fluid point and then sweeps in the positive direction of $v_1$. When it reaches the most right fluid point, it continues at the left most fluid point of the upward row in the positive direction of $v_2$. Repeat the rightward and upward sweep until it reaches the top-right fluid point. One can imagine that a 3D porous structure can be created from this 2D porous structure by duplicating 2D slice in the positive direction of the $x_3$ coordinate axis. Considering the group of discrete velocity $v_j^{(k)}=(v_1^{(k)}>0,v_2^{(k)}>0,v_3^{(k)}>0)$ in that 3D porous sample, the volume path of sweep can be developed from the above plane path of sweep by continuing from the bottom-left fluid point of the next plane in the positive direction of $v_3$, and so forth. Paths of sweep for the other three (seven) groups of discrete velocity in 2D (3D) simulations can be deducted in an analogous manner. Distribution function of a fluid point is computed from these of the neighbour points in the upwind stencil, which have been already calculated either in some early stages of the current sweep (if the neighbours are the fluid points) or in the step~\ref{it:BC} of the previous iteration (if the neighbours are boundary points). It should be noted that each solid corner point has either 2 or 3 outer normal unit vectors  $\mathbf{n}_j$ directing to its neighbour fluid points. If the numerical stencil includes a solid corner point that links to the central fluid point in the $x_j$-direction, diffuse-specular reflection at that corner point must be calculated with the normal vector $\mathbf{n}_j$ by Eq.~\eqref{eq:diffusive-specular_BC}. For example, considering the numerical stencil of the fluid point on the right of the corner point in Fig.~\ref{fig:halo_sweep_stencil}, diffuse-specular reflection at that corner point must be calculated with the normal vector $\mathbf{n}_2=(1,0)$.
	
	\item \label{it:BC} Applying the \textit{boundary conditions}, i.e. Eq.~\eqref{eq:diffusive-specular_BC} on solid surfaces, and updating the boundary conditions, i.e. ~Eqs.~\eqref{eq:periodic_BC} and~\eqref{eq:symmetric_BC} on the outer faces of the porous sample.

	\item \label{it:macroscopic} The \textit{macroscopic flow fields} obtained from Eq.~\eqref{eq:perturbed_quantities} and the gas number density on the solid surface from Eq.~\eqref{eq:wall_density} are updated by the
	quadrature rule associated with the adopted velocity grid. 
	
\end{enumerate}
Only the resulting data for step~\ref{it:BC} and step~\ref{it:macroscopic} are needed for the next iteration, thus we do not need to store the solved distribution function $h$ in the bulk region as required in the time-marching schemes. The two supplementary steps are performed at every 100 iterations: 
\begin{enumerate}[label=(\roman*),resume]
	\item \textit{Permeability} $k$ given by Eq.~\eqref{eq:k_Gp} is approximated by the trapezoidal rule applied on a transverse cross section. 
	\item The values of permeability obtained at the current iteration $k^{(l)}$ and the previous $100$ iterations $k^{(l-100)}$ are used to assess \textit{convergence}, i.e.
	\begin{equation}
	\begin{aligned}
	Err(l)=\dfrac{1}{100}\left|\dfrac{k^{(l)}-k^{(l-100)}}{k^{(l)}} \right| < 10^{-8}.      
	\end{aligned}
	\label{eq:convergence_criterion}
	\end{equation}
\end{enumerate}

\begin{algorithm}
	
	\SetKwProg{COpenMPParallel}{\textcolor{blue}{\#pragma\enspace omp\enspace parallel}}{}{}
	\SetKwProg{COpenMPDo}{\textcolor{blue}{\#pragma\enspace omp\enspace for}}{}{}
	\SetKwProg{COpenMPSingle}{\textcolor{blue}{\#pragma\enspace omp\enspace single}}{}{}
	\SetKwFunction{OMPDo}{DO}
	\SetKwFunction{FindH}{FindH}
	\SetKwFunction{FindBC}{FindBC}
	\SetKwFunction{FindMacro}{FindMacro}
	\SetKwInOut{Input}{data}\SetKwInOut{Output}{results}
	\Input{$h^{BC}$ and $\varrho,\mathbf{u},\tau$ obtained from the previous iteration or the initial conditions}
	\Output{updated $h^{BC}$ and $\varrho,\mathbf{u},\tau$}
	\BlankLine
	\COpenMPParallel{}{
		\HiLi \COpenMPSingle{}{
			\HiLi \textcolor{red}{MPI\_ISEND}(sending\_buffers) \tcp*[r]{nonblocking send}
			\HiLi \textcolor{red}{MPI\_IRECV}(receving\_buffers) \tcp*[r]{nonblocking receive}
		}
		\COpenMPDo{}{
			\ForAll{discrete velocity $\mathbf{v}^{(k)}$}{
				\For(\tcp*[f]{Raster scan on all fluid points (forward substitution), see fig.~\ref{fig:halo_sweep_stencil}}){$i = 1$ \KwTo $N_{x}$}{
					$h_{k,i}$ $\leftarrow$ \FindH{$k,i$}
					\tcp*[r]{Advection \& collision Eqs.\eqref{eq:linearised_Boltzmann_equation},\eqref{eq:numerical stencil}}
				}
			}
		}
		\HiLi \COpenMPSingle{}{
			\HiLi \textcolor{red}{MPI\_WAITALL} \tcp*[r]{wait for send \& receive done}
		}
		\HiLi \COpenMPDo{}{
			\HiLi \ForAll{external halo (ghost points) and internal halo}{
				\HiLi $h^{ext\_halo}$ $\leftarrow$ receiving\_buffers \tcp*[r]{buffer unpack}
				\HiLi sending\_buffers $\leftarrow$ $h^{int\_halo}$ \tcp*[r]{buffer pack}
			}
		}
		\COpenMPDo{}{
			\ForAll{boundary points $i$}{
				$h_{i}^{BC}$ $\leftarrow$ \FindBC{$i$} 
				\tcp*[r]{Boundary conditions Eqs.\eqref{eq:diffusive-specular_BC},\eqref{eq:periodic_BC},\eqref{eq:symmetric_BC}}
			}
		}
		\COpenMPDo{}{
			\ForAll{fluid points $i$}{
				$\varrho_i,\mathbf{u}_i,\tau_i$ $\leftarrow$ \FindMacro{$h_{i}$} 
				\tcp*[r]{Macroscopic parameters Eqs.\eqref{eq:perturbed_quantities}}
			}
		}
	}
	\caption{Two-level parallel MPI/OpenMP pseudo-code for one iteration.
 \textcolor{red}{MPI routines} and \textcolor{blue}{OpenMP compiler directives} are in red and blue colors, respectively. By removing highlighted lines, which represent work on data transfers between spatial subdomains, one obtains pure OpenMP pseudo-code. By discarding blue lines, one obtains pure MPI pseudo-code with spatial domain decomposition. The serial pseudo-code corresponds to un-highlighted black lines.} \label{algo_hybrid}
	\end{algorithm}
	

\subsection{Two-level parallel implementation using MPI and OpenMP}
%
%
%


The two-level parallel approach as illustrated in the Algorithm~\ref{algo_hybrid} is a natural option when the discretizations in both the spatial and velocity spaces are required.
At the upper level, the 3D/2D spatial domain $\Omega_x$ is decomposed into multiple cuboid/rectangular subdomains, and each subdomain is assigned to one MPI process.
At the lower level, i.e. within each subdomain, OpenMP is employed to parallelize 
the major steps described in Sec.~\ref{sec:serial_implementation}.
Moreover, additional steps are introduced to deal with data transfer between subdomains, e.g. buffer packing/unpaking are also parallelized with OpenMP.  
In the rest of this paper, multi-level MPI/OpenMP parallelization will be denoted by MPI $c_1\times c_2\times c_3$ OMP $c_4$. Here $c_1, c_2, c_3$ are the number of subdomain divisions along the $x_1, x_2, x_3$ coordinate axes, respectively, while $c_4$ is the number of OpenMP threads allocated to each subdomain. Therefore, the total number of subdomains, i.e. the MPI processes, is $c_1\times c_2\times c_3$ and the total number of adopted cores is $c_1\times c_2\times c_3 \times c_4$. Pure MPI or pure OpenMP parallelizations correspond to $c_4=1$ or $c_1=c_2=c_3=1$, respectively.  

MPI communications occur only between the spatial subdomains, see Fig.~\ref{fig:halo_sweep_stencil}. To facilitate the overlapping of computation and MPI communication, we use the classical ghost points approach and the non-blocking version of the MPI send/receive subroutines.
To be more specific, two layers of additional grid points, called ghost points or external halo, are padded around each computational subdomain, resulting an extended subdomain. As a result, the fluid points near the communication boundaries, called internal halo, can be treated normally like the inner fluid points, although the numerical stencils around the internal halo may occupy points beyond the communication boundaries.
Additional sending (receiving) buffers are allocated to store the distribution function of the out-going (in-coming) discrete velocities at each subdomain boundary.
The non-blocking MPI send/receive subroutines and the MPI wait subroutine are called prior and posterior, respectively, to the processing of the bulk subdomains.
Then, the out-going (in-coming) distribution functions at the internal halo (external halo) are packed (unpacked) to (from) the sending (receiving) buffers.
By using the two-level MPI/OpenMP parallel approach, the total number of MPI communication function calls is effectively reduced compared with the pure MPI approach, so that the communication congestion is avoided and the communication time is reduced so as to realize MPI level optimization, in which the performance of the program is improved from two aspects, MPI level optimization and algorithmic level optimization.


As stated in Sec.~\ref{sec:intro},  the resulting parallel solver is not equivalent to the original serial solver in terms of the convergence history, because the grid points near the communication boundaries use the last iteration value of the upwind grid points. In the following Section, we will investigate how and to what extend the convergence rate is affected by the spatial domain decomposition, and even more importantly, whether the final permeability converges to the prediction of serial solver.



%
\section{Convergence properties of the kinetic solvers}\label{sec:Deterioration}



Both the 2D and 3D MPI/OpenMP parallel implementations of the current algorithm are compared against the OpenMP parallelization (without domain decomposition) in terms of the convergence rate and converged permeability. It is noted that the OpenMP parallelization has been validated for 2D flow through square array of circular cylinders in our previous work~\cite{Wul:2017}. The term associated with the pertubed temperature $\tau$ in Eq.~\eqref{eq:Discretised_linearised_Boltzmann_equations} has no influence on permeability in our test on isothermal flow thus it is neglected. The solvers are written in Fortran and compiled by the Cray Programming Environment (version 5.2.82), and run on the ARCHER super-computing system (${118080}$ processing cores in total), the UK national academic HPC facility. Each compute node of the ARCHER contains two 12-core E5-2697 v2 (Ivy Bridge) series processors and 64 GB memory. 

\begin{figure*}[t]
\centering
\begin{subfigure}[t]{0.375\textwidth}
\centering
\includegraphics[width=0.87\textwidth]{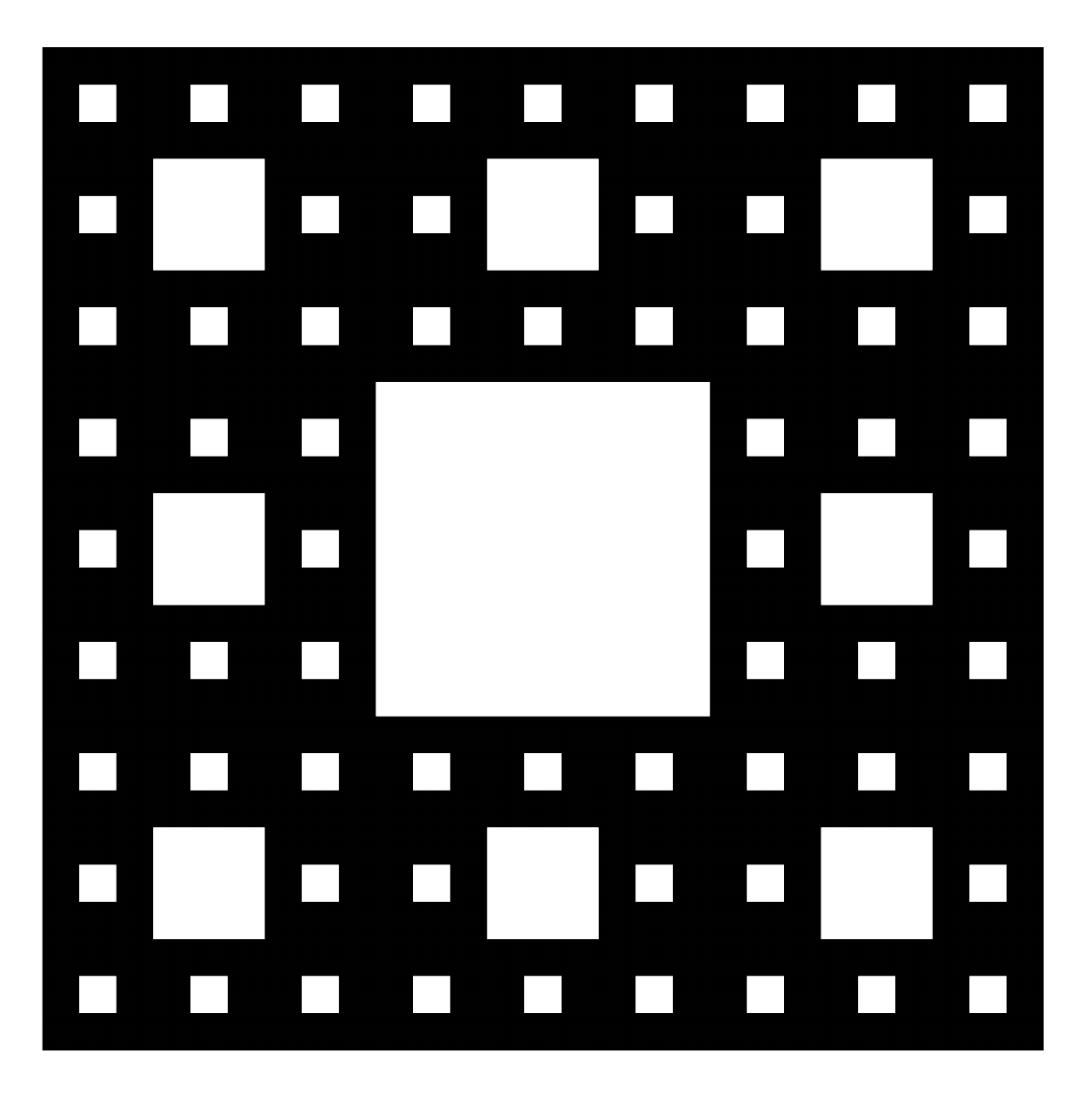}
\caption{The level-3 Sierpinski carpet}
\label{fig:carpet}
\end{subfigure}%
~   
\begin{subfigure}[t]{0.625\textwidth}
\centering
\includegraphics[width=\textwidth]{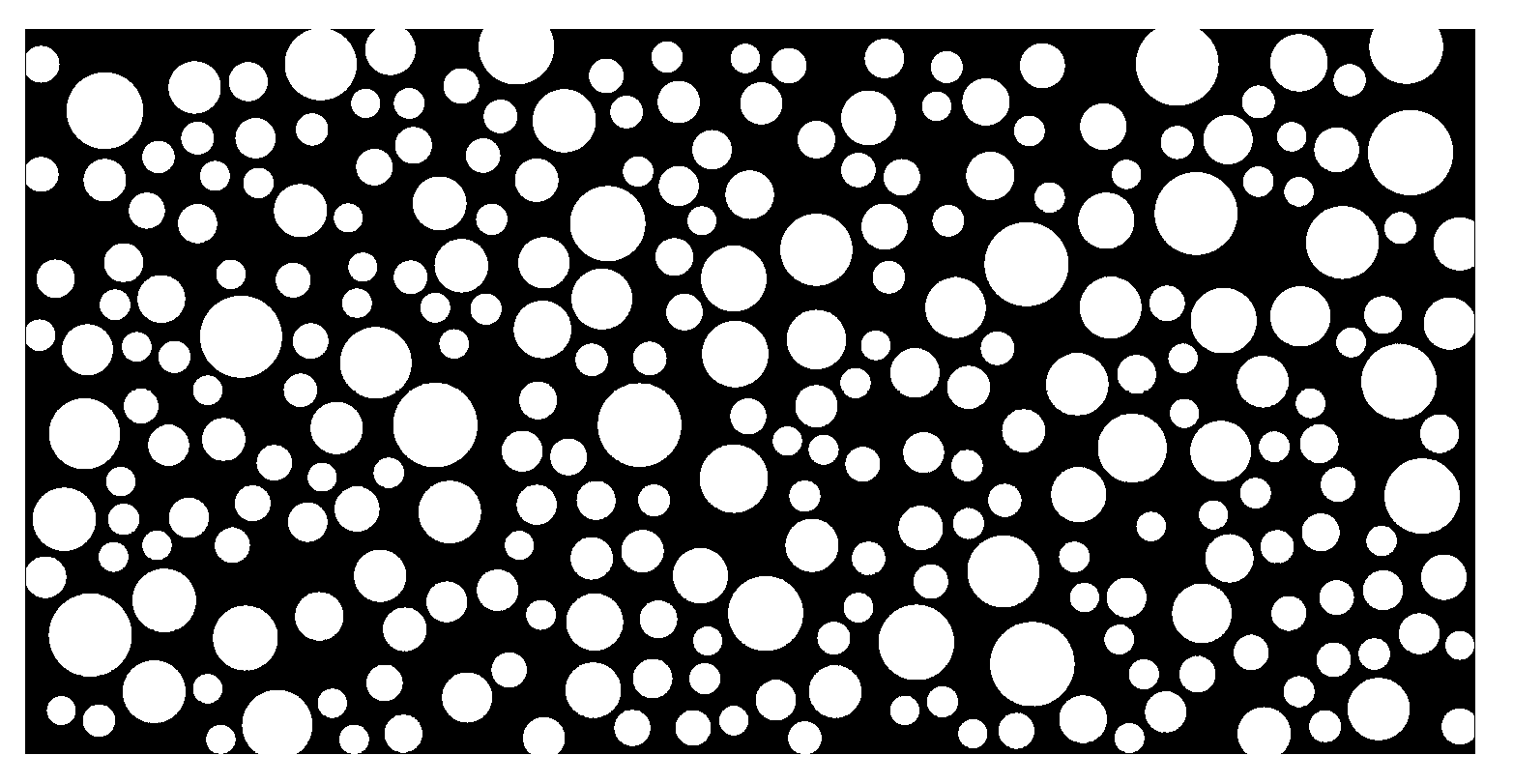}
\caption{Circular cylinders of random size and position}
\label{fig:random_cylinders}
\end{subfigure}%
\caption{2D digital models of porous media, in which the white and black regions represent the matrix and voids, respectively. The spatial grid size $N_x$ and porosity $\epsilon$ are (a) $N_x=540\times 540$, $\epsilon=0.70$; and (b) $N_x=3000\times 1500$, $\epsilon=0.60$.} 
\label{fig:carpet_cylinders}
\end{figure*}

\subsection{2D flows}
\begin{figure*}[t]
	\centering
	\begin{subfigure}[t]{0.5\textwidth}
		\centering
		\includegraphics[width=\textwidth]{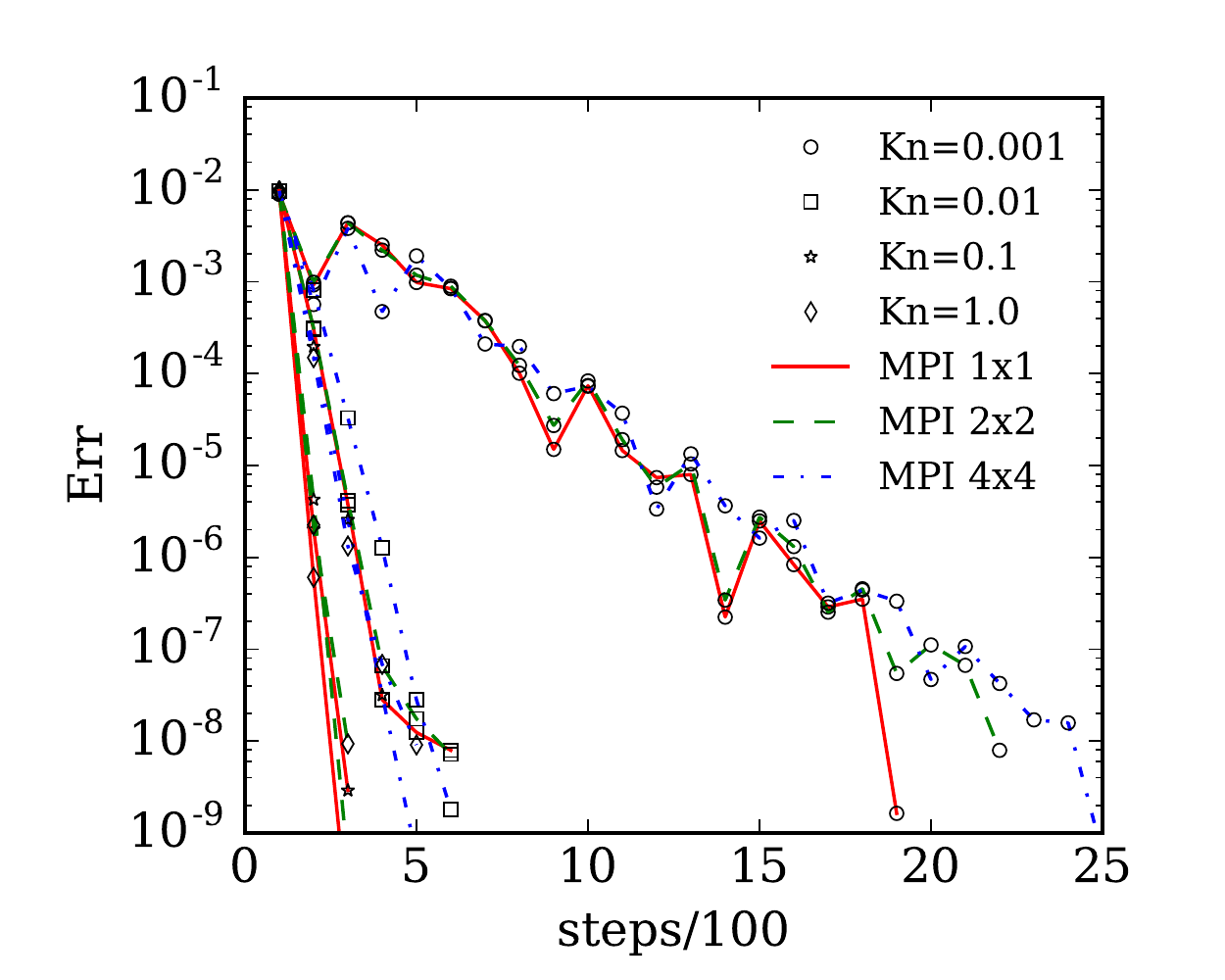}
		\caption{Sierpinski carpet}
		\label{fig:logErr_carpet}
	\end{subfigure}%
	~   
	\begin{subfigure}[t]{0.5\textwidth}
		\centering
		\includegraphics[width=\textwidth]{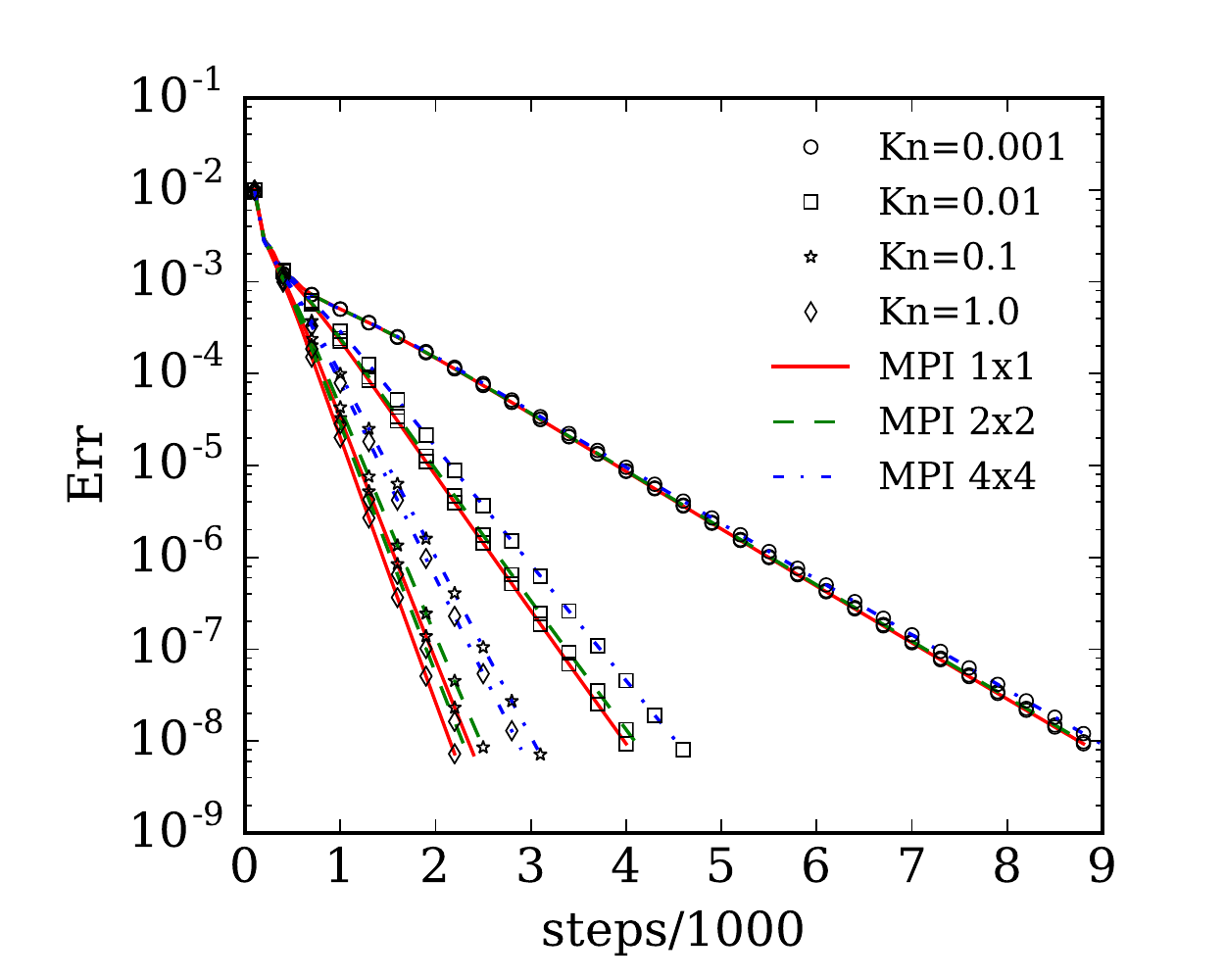}
		\caption{2D random cylinders}
		\label{fig:logErr_cylinders}
	\end{subfigure}%
	
	\caption{ Convergence history of the permeabilities given by Eq.~\eqref{eq:convergence_criterion} at different Knudsen numbers $Kn$ with different spatial domain decomposition for the 2D models of porous media as shown in Fig.~\ref{fig:carpet_cylinders}. 
	}
	\label{fig:logErr_carpet_cylinders}
\end{figure*}

We first consider a regular model porous medium, i.e. the Sierpinski carpet, which is a famous plane fractal and can be constructed through recursion.
The construction begins with a square. The square is cut into 9 congruent sub-squares with a 3-by-3 grid, and the central sub-square is then removed. The same procedure is applied recursively to the remaining 8 sub-squares. We consider a three-level construction with a resolution of $540\times540$ pixels, see Fig.~\ref{fig:carpet_cylinders}(a).
The Knudsen number $Kn$ is associated with the reference length $L=540$ pixels taken from size of the sample in the $x_1$ direction, while the effective Knudsen number $Kn^*=34.40Kn$ is estimated by Eq.~\eqref{eq:effective_Kn} with $\epsilon=0.70$ and $k_\infty=4.930\times10^{-5}$. 
We consider $Kn$ of $0.001, 0.01, 0.1, 1$, so the flow ranges from the slip to free-molecular flow regimes.
The discrete velocity grids for the case of $Kn=1$ are 
$24\times24$, and for the others are $8\times8$.
For each Kn, we run the parallel solver with $2\times2$ and  $4\times4$ even decomposition.
The numbers of iterations as well as the wall clock time to reach the convergence criterion Eq.~\eqref{eq:convergence_criterion} for each kind of parallel decomposition are tabulated in Table~\ref{tab:carpet}.
In the bottom row of the table, we also present the converged permeabilities obtained by the pure OpenMP computations. It is noted that pure OpenMP option (MPI $1\times 1$ OMP 12) predicts the identical results as the serial version (MPI $1\times 1$ OMP 1) in terms of convergence history and converged permeabilities where no domain decomposition is applied.  
The permeabilities predicted by the multi-level parallel are not shown here, because they match those of the pure OpenMP option, with the maximum relative deviation being $0.000203\%$. The convergence history of the permeabilities at different Knudsen numbers and domain decompositions are shown in Fig.~\ref{fig:logErr_carpet_cylinders}(a). Both Table~\ref{tab:carpet} and Figure~\ref{fig:logErr_carpet_cylinders}(a) show that the deterioration in convergence rate generally becomes worse with increasing spatial subdomains and Knudsen number. The number of iterations has risen by $32\%$ and $67\%$ for $Kn=0.001$ and $Kn=1$, respectively, when $16$ subdomains are adopted. 
The slow convergence rate near the continuum regime for the classical DVM, without domain decomposition, has been confirmed by the Fourier stability analysis, where the spectral radius is found to be equal to unity~\cite{Valougeorgis:2003}.

\begin{table}[t]
	\centering
	\caption{2D porous model of the Sierpinski carpet [Fig.~\ref{fig:carpet_cylinders}(a)]: the wall clock time (in second)
		and the number of iterations (in curve parentheses) at various $Kn$ with different spatial domain $\Omega_{x}$-decomposition. The velocity grid size $N_{v}=24^{2}$
		and $N_{v}=8^{2}$ are used for $Kn=1$ and $Kn=0.1,0.01,0.001$,
		respectively.}
	\begin{tabular}{ccccccccc}
		\hline 
		$Kn \ (Kn^\ast=34Kn)$ & \multicolumn{2}{c}{$0.001$} & \multicolumn{2}{c}{$0.01$} & \multicolumn{2}{c}{$0.1$} & \multicolumn{2}{c}{$1$}\tabularnewline
		\hline 
		MPI $1\times1$ OMP $12$ & $53.39s$ & $(1900)$ & $16.86s$ & $(600)$ & $8.43s$ & $(300)$ & $94.21s$ & $(300)$\tabularnewline
		MPI $2\times2$ OMP $12$ & $13.50s$ & $(2200)$ & $3.69s$ & $(600)$ & $1.84s$ & $(300)$ & $18.60s$ & $(300)$\tabularnewline
		MPI $4\times4$ OMP $12$ & $4.23s$ & $(2500)$ & $1.02s$ & $(600)$ & $0.87s$ & $(500)$ & $11.20s$ & $(500)$\tabularnewline
		\hline 
		$k$ & \multicolumn{2}{c}{$4.930\times10^{-5}$} & \multicolumn{2}{c}{$1.355\times10^{-4}$} & \multicolumn{2}{c}{$1.143\times10^{-3}$} & \multicolumn{2}{c}{$1.490\times10^{-2}$}\tabularnewline
		\hline 
	\end{tabular}
	\label{tab:carpet}
\end{table}

Now we considered a more complex 2D porous media composed by circular cylinders with random locations and sizes, as shown in Fig.~\ref{fig:carpet_cylinders}(b), where the porosity is 0.6. The spatial grid size is $3000\times1500$, and $Kn=0.001, 0.01, 0.1, 1$. The velocity grids $N_v$ used for the smaller $Kn$  (0.001 and 0.01) and larger $Kn$ (0.1 and 1) cases are $8\times8$ and $24\times24$, respectively. 
The Knudsen number $Kn$ is defined with the reference length $L=3000$ pixels,
 while the effective Knudsen number $Kn^*=64.52Kn$ is
estimated by Eq.~\eqref{eq:effective_Kn} with $k_\infty=1.201\times10^{-5}$. 
Similar to the previous Sierpinski carpet case, we show that the number of iterations to reach the converged permeabilities as presented in Table~\ref{tab:cylinders} and the convergence histories of permeability as shown in Fig.~\ref{fig:logErr_carpet_cylinders}(b). 
The deterioration in convergence rate due to the spatial domain decomposition is found to be mitigated by increasing complexity in porous structure, especially at small Knudsen numbers. The increase of the total iterations is only $2\%$ for $Kn=0.001$ and $32\%$ for $Kn=1$, when $16$ subdomains are used.
The maximum deviation of the permeabilities predicted by the parallel solver from the corresponding serial ones is 0.024\%. Comparing Fig.~\ref{fig:logErr_carpet_cylinders}(b) with Fig.~\ref{fig:logErr_carpet_cylinders}(a), we can find the convergence history of the random circular cylinders case is much smoother then the regular Sierpinski carpet case. This can be explained by the more pronounced mixing effect caused by the complex flow passages and gas-surface interactions.


\begin{table}[t]
	\centering
	\caption{The 2D porous model of random cylinders [Fig.~\ref{fig:carpet_cylinders}(b)]: the wall clock time (in second)
		and the number of iterations (in curve parentheses) at various Knudsen
		number $Kn$ with different spatial domain $\Omega_{x}$-decomposition.
		The velocity grid size $N_{v}=24^{2}$
		and $N_{v}=8^{2}$ are used for $Kn=1,0.1$ and $Kn=0.01,0.001$,
		respectively.}
	\begin{tabular}{ccccccccc}
		\hline 
		$Kn (Kn^\ast=65Kn)$ & \multicolumn{2}{c}{$0.001$} & \multicolumn{2}{c}{$0.01$} & \multicolumn{2}{c}{$0.1$} & \multicolumn{2}{c}{$1$}\tabularnewline
		\hline 
		MPI $1\times1$ OMP $12$ & $2719s$ & $(8800)$ & $1236s$ & $(4000)$ & $8208s$ & $(2400)$ & $7524s$ & $(2200)$\tabularnewline
		MPI $2\times2$ OMP $12$ & $730s$ & $(8800)$ & $344s$ & $(4100)$ & $2380s$ & $(2500)$ & $2160s$ & $(2300)$\tabularnewline
		MPI $4\times4$ OMP $12$ & $234s$ & $(9000)$ & $120s$ & $(4600)$ & $791s$ & $(3100)$ & $740s$ & $(2900)$\tabularnewline
		\hline 
		$k$ & \multicolumn{2}{c}{$1.201\times10^{-5}$} & \multicolumn{2}{c}{$4.524\times10^{-4}$} & \multicolumn{2}{c}{$3.758\times10^{-3}$} & \multicolumn{2}{c}{$3.662\times10^{-2}$}\tabularnewline
		\hline 
	\end{tabular}
	\label{tab:cylinders}
\end{table}


\subsection{3D flows}
Similar verifications are also conducted for our 3D parallel implementation of the algorithm on the flows through a simple cubic array of spheres and the randomly packed spheres, which are often used as model porous media in analytical, computational and experimental studies~\cite{Sangani:1982,Hill:2001}. 

\begin{figure}[thbp]
	\centering
	\begin{subfigure}[t]{0.5\textwidth}
		\centering
		\includegraphics[width=\textwidth]{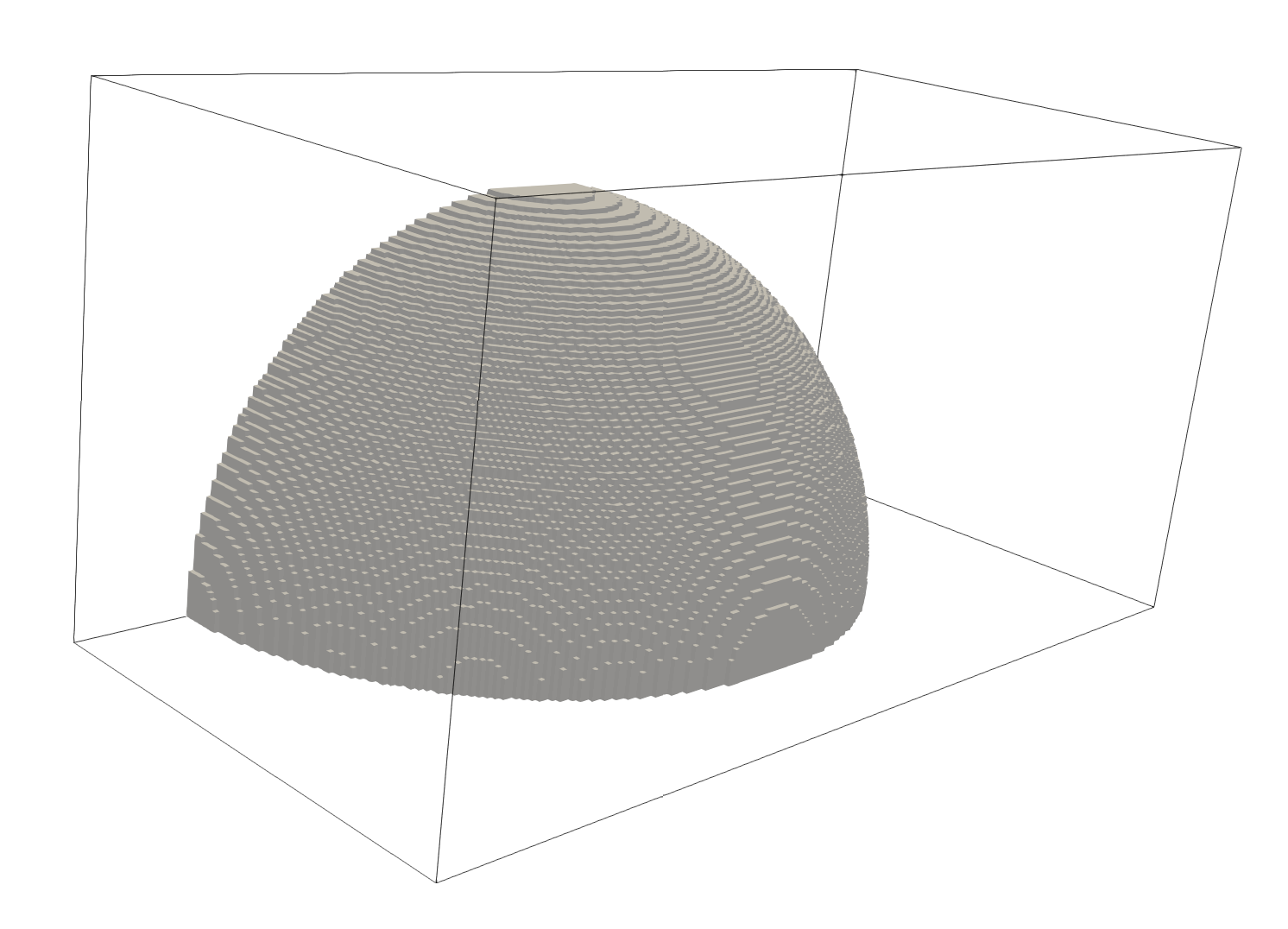}
		\caption{Cubic sphere packing}
		\label{fig:singlesphere}
	\end{subfigure}%
	~   
	\begin{subfigure}[t]{0.5\textwidth}
		\centering
		\includegraphics[width=0.8\textwidth]{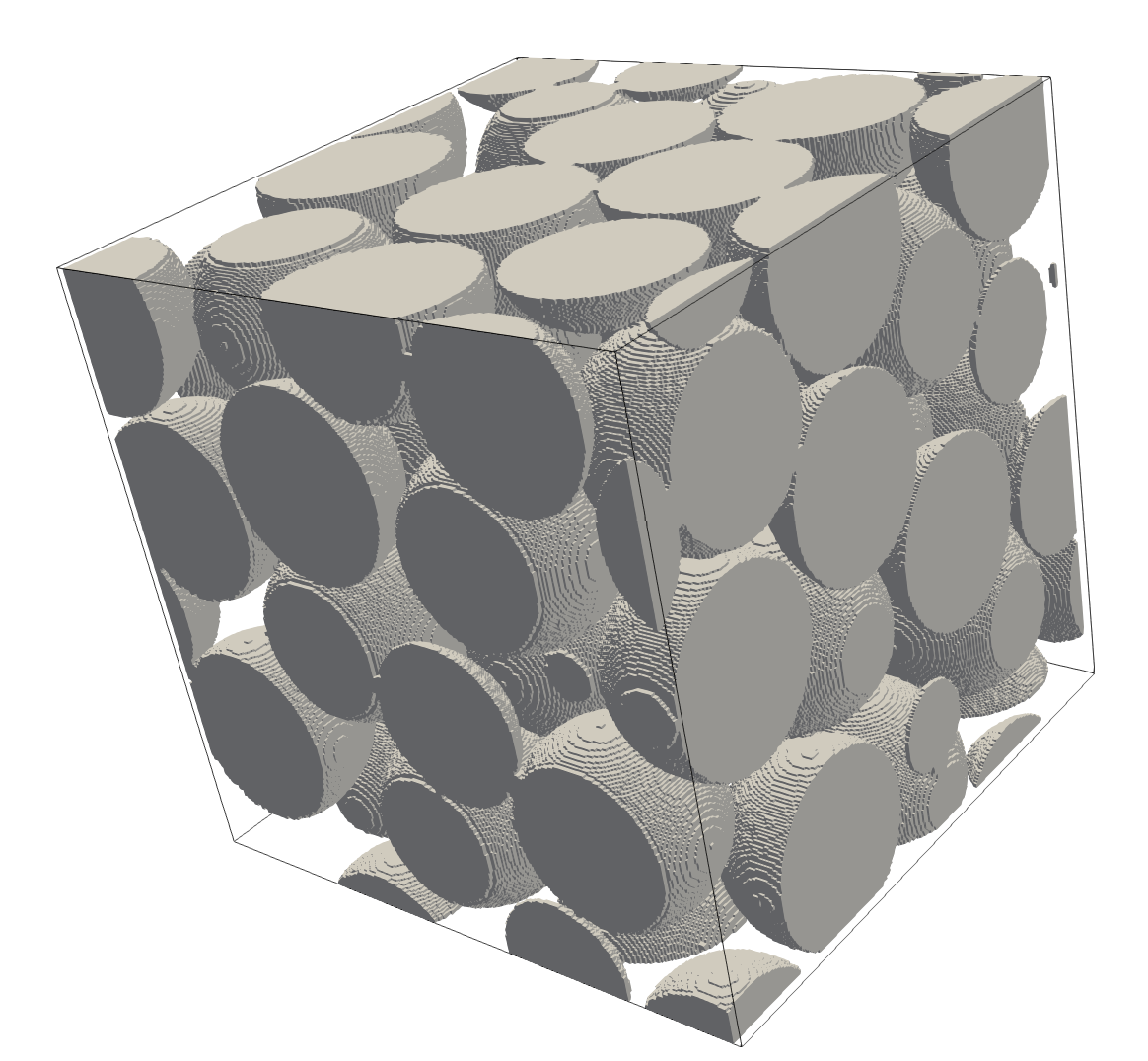}
		\caption{Irregular sphere packing}
		\label{fig:beadpack}
	\end{subfigure}%
	
	\begin{subfigure}[t]{0.5\textwidth}
	\centering
	\includegraphics[width=\textwidth]{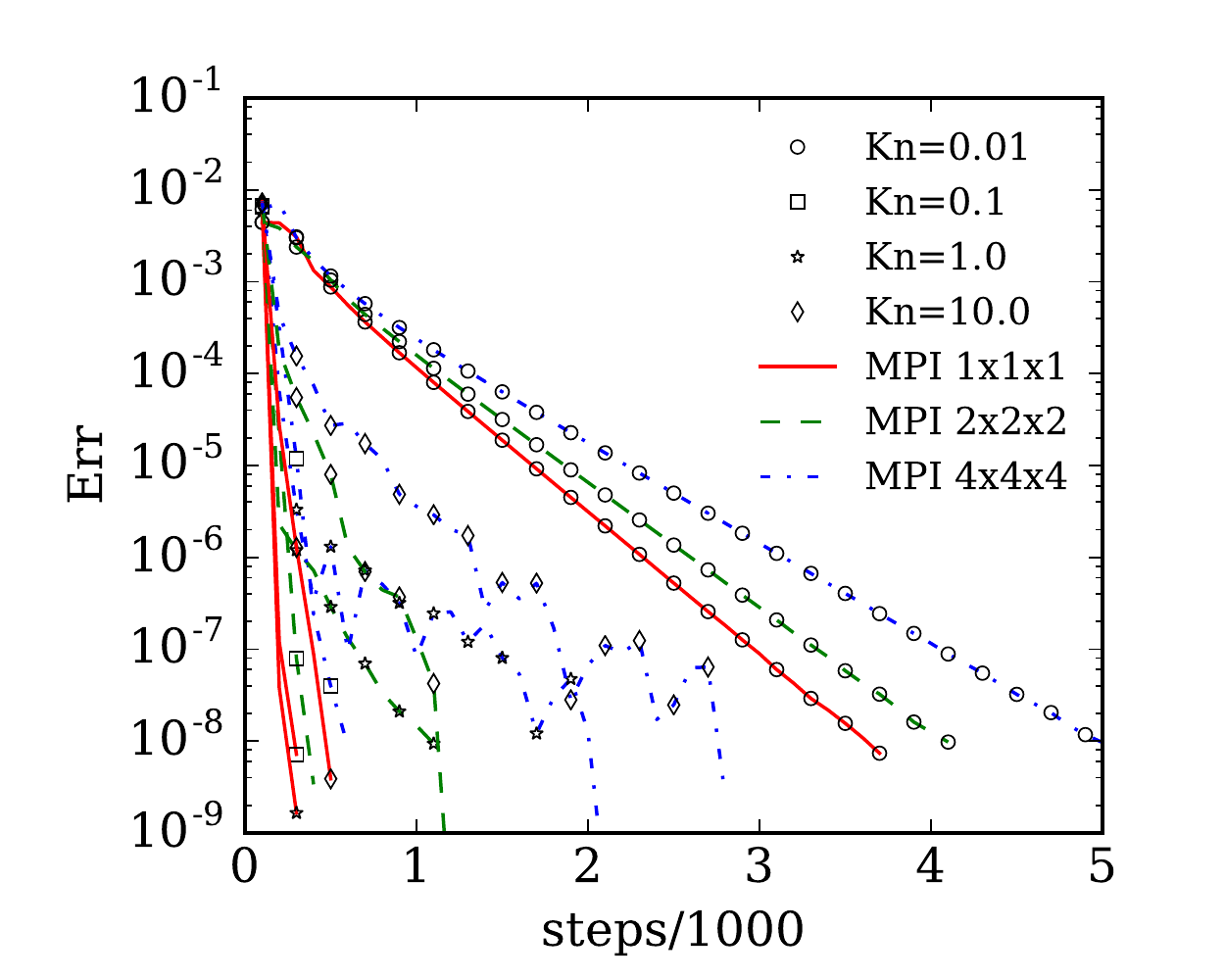}
	\caption{3D cubic sphere packing}
	\label{fig:logErr_singlesphere}
	\end{subfigure}%
	~   
	\begin{subfigure}[t]{0.5\textwidth}
	\centering
	\includegraphics[width=\textwidth]{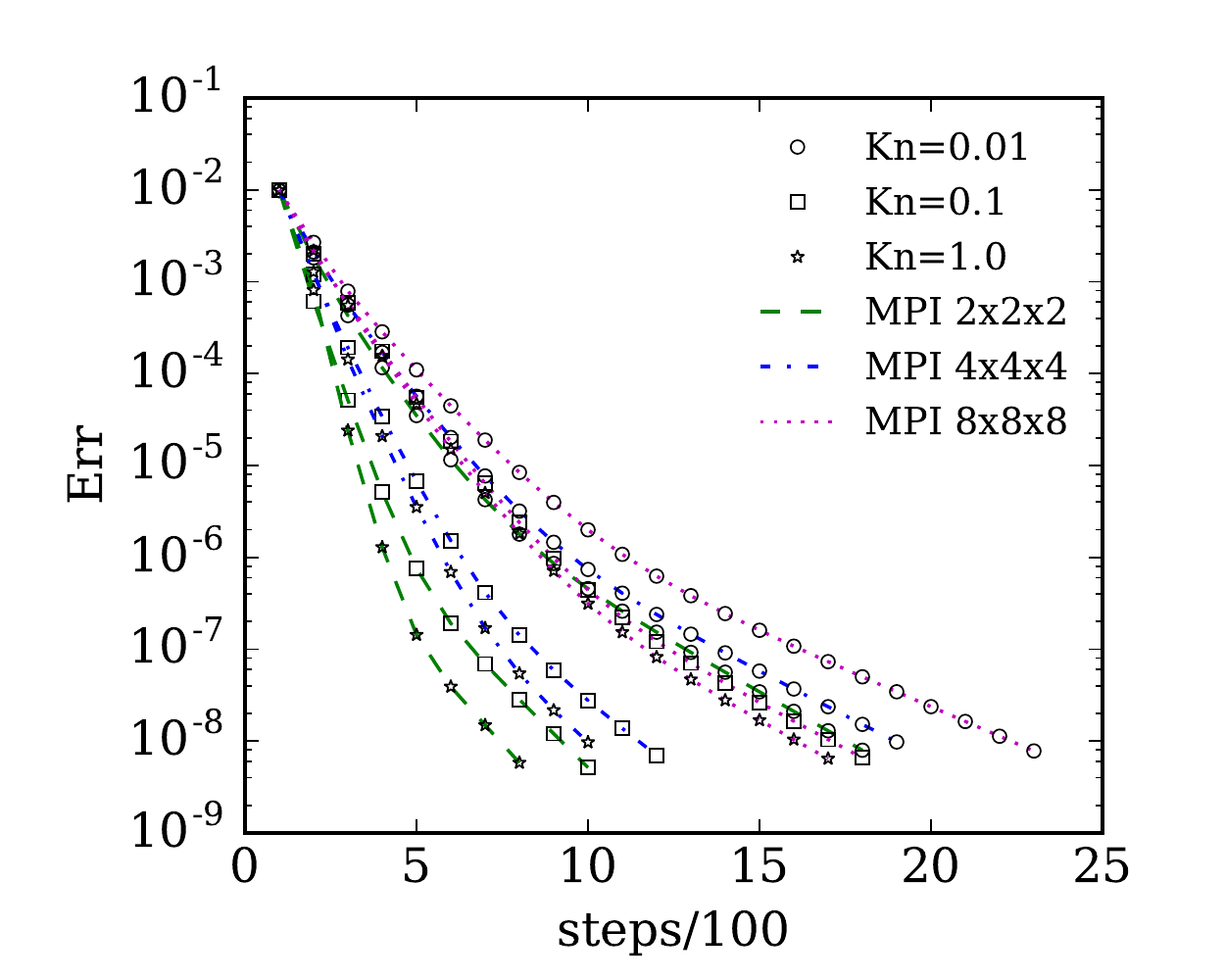}
	\caption{3D irregular sphere packing }
	\label{fig:logErr_beadpack}
	\end{subfigure}%

	\caption{The computational domains for 3D models of porous media made by sphere (in grey color) packing and the convergence history of the permeabilities given by Eq.~\eqref{eq:convergence_criterion} at different Knudsen numbers $Kn$ with different spatial domain decomposition. The spatial grid size $N_x$ and porosity $\epsilon$  
	are (a) $N_x=200\times100\times100$, $\epsilon=0.75$,  (b) $N_x=308\times300\times300$, $\epsilon=0.38$.}
	\label{fig:singlesphere_beadpack}
\end{figure}

In the first case, the unit cell is a cubic box with a sphere located at its centre and repeated itself in the 3D space. Due to symmetry and periodicity of the configuration, we simulate only a quadrant of the unit cell as shown in Fig.~\ref{fig:singlesphere_beadpack}(a) with the spatial grid points of $200\times100\times100$. The Knudsen number $Kn$ is associated with the reference length $L=200$ voxels, taken from the sample size in the $x_1$ direction, while the effective Knudsen number $Kn^*=4.14Kn$ is estimated by Eq.~\eqref{eq:effective_Kn} with $\epsilon=0.75$ and $k_\infty=1.722\times10^{-2}$. The number of iterations and the converged values of permeability are listed in Table~\ref{tab:singlesphere}, and the convergence histories are also presented in Fig.~\ref{fig:singlesphere_beadpack}(c). The permeabilities predicted by the parallel solver deviate from the serial solver's results up to $0.11\%$.  
Similar to the 2D porous models, the degeneration of convergence rate also increases with number of spatial subdomains and Knudsen number. The number of iterations is increased by $35\%$ and $460\%$ for $Kn=0.01$ and $Kn=10$, respectively, when $64$ subdomains are applied.
We can also observe from Fig.~\ref{fig:singlesphere_beadpack}(c) that the parallel solver's convergence history is not as smooth as the serial solver's at the high-$Kn$ cases. One possible reason is that at a high $Kn$, some particles are altered their distribution function accidentally due to one-interation-lag in information transfer (at the boundaries of subdomain) instead of moving freely between the inlet and the outlet (as in the serial solver). These alternations then influence back to the inlet or outlet through periodic boundary conditions.


%

The second case is a cube filled with randomly packed spheres as illustrated in Fig.~\ref{fig:singlesphere_beadpack}(b). 
Note that we extend four fluid layers at the inlet and outlet of the original geometry to make periodic boundary conditions applicable. 
The spatial grid size is $308\times300\times300$ including the extended fluid layers. The Knudsen number $Kn$ is associated with the reference length $L=308$ voxels,
 while the effective Knudsen number $Kn^*=34.41Kn$ is estimated by Eq.~\eqref{eq:effective_Kn} with $\epsilon=0.38$ and $k_\infty=1.260\times10^{-4}$. $Kn$ of $0.01, 0.1$ and $1$ are considered, and the velocity grid points of $8\times8\times8$ are deployed. The number of iterations together with the converged values of the permeabilities are listed in Table~\ref{tab:beadpack}.
From Fig.~\ref{fig:singlesphere_beadpack}(d), it can be seen that the convergence history of the irregular sphere packing case is smooth even at high $Kn$, unlike the simple cubic array of spheres case. Table~\ref{tab:beadpack} shows that the slowing down of convergence rate in the parallel solver is not as significant as the simple cubic array of spheres case. 
The number of iterations increases by only $28\%$ and $113\%$ for $Kn=0.01$ and $Kn=1$, respectively, when the number of subdomains increases from $8$ to $512$.
These observations in 3D porous media confirm our finding in the 2D cases that degeneration in convergence rate due to spatial domain decomposition is alleviated by complex porous media.


\begin{table}[thbp]
	\centering
	\caption{The 3D porous model of cubic sphere packing (Fig.~\ref{fig:singlesphere}): the wall clock time (in
		second) and the number of iterations (in curve parentheses) at various
		Knudsen number $Kn$ with different spatial domain $\Omega_{x}$-decomposition. The velocity grid size $N_{v}=16^{3}$
		and $N_{v}=8^{3}$ are used for $Kn=10,1$ and $Kn=0.1,0.01$, respectively. }
	\begin{tabular}{ccccccccc}
		\hline 
		$Kn \ (Kn^\ast=4.1Kn)$ & \multicolumn{2}{c}{$0.01$} & \multicolumn{2}{c}{$0.1$} & \multicolumn{2}{c}{$1$} & \multicolumn{2}{c}{$10$}\tabularnewline
		\hline 
		MPI $1\times1\times1$ OMP $12$ & $10663s$ & $(3700)$ & $864.6s$ & $(300)$ & $ 7830s$ & $(300) $ & $13050s$ & $(500) $\tabularnewline
		MPI $2\times2\times2$ OMP $12$ & $1974s $ & $(4100)$ & $192.6s$ & $(400)$ & $4314s$ & $(1100)$ & $4706s$ & $(1200)$\tabularnewline
		MPI $4\times4\times4$ OMP $12$ & $377.4s$ & $(5000)$ & $45.28s $ & $(600)$ & $1397s$ & $(2100)$ & $1865$ & $(2800)$\tabularnewline
		\hline 
		$k$ & \multicolumn{2}{c}{$1.722\times10^{-2}$} & \multicolumn{2}{c}{$3.862\times10^{-2}$} & \multicolumn{2}{c}{$2.836\times10^{-1}$} & \multicolumn{2}{c}{$3.375\times10^{0}$}\tabularnewline
		\hline 
	\end{tabular}
	\label{tab:singlesphere}
\end{table}

\begin{table}[thbp]
	\centering
	\caption{The 3D porous model of irregular sphere packing [Fig.~\ref{fig:singlesphere_beadpack}(b)]: the wall clock time
		(in second) and the number of iterations (in curve parentheses) at various
		Knudsen number $Kn$ with different spatial domain $\Omega_{x}$-decomposition. The data for pure OpenMP option is not available due to memory limit of one compute node.}
	\begin{tabular}{ccccccc}
		\hline 
		$Kn\ (Kn^\ast=34Kn)$ & \multicolumn{2}{c}{$0.01$} & \multicolumn{2}{c}{$0.1$} & \multicolumn{2}{c}{$1$}\tabularnewline
		\hline 
		MPI $1\times1\times1$ OMP $12$ & \multicolumn{6}{c}{n/a}\tabularnewline
		MPI $2\times2\times2$ OMP $12$ & $8057s$ & $(1800)$ & $4475s$ & $(1000)$ & $3580s$ & $(800)$\tabularnewline
		MPI $4\times4\times4$ OMP $12$ & $1318s$ & $(1900)$ & $831.8s$ & $(1200)$ & $692.9s$ & $(1000)$\tabularnewline
		MPI $8\times8\times8$ OMP $12$ & $286.4s$ & $(2300)$ & $224.1s$ & $(1800)$ & $211.7s$ & $(1700)$\tabularnewline
		\hline 
		$k$ & \multicolumn{2}{c}{$1.260\times10^{-4}$} & \multicolumn{2}{c}{$7.823\times10^{-4}$} & \multicolumn{2}{c}{$7.361\times10^{-3}$}\tabularnewline
		\hline 
	\end{tabular}
	\label{tab:beadpack}
\end{table}


\begin{figure}[th]
	\centering
	\begin{subfigure}[t]{0.45\textwidth}
		\centering
		\includegraphics[width=\textwidth]{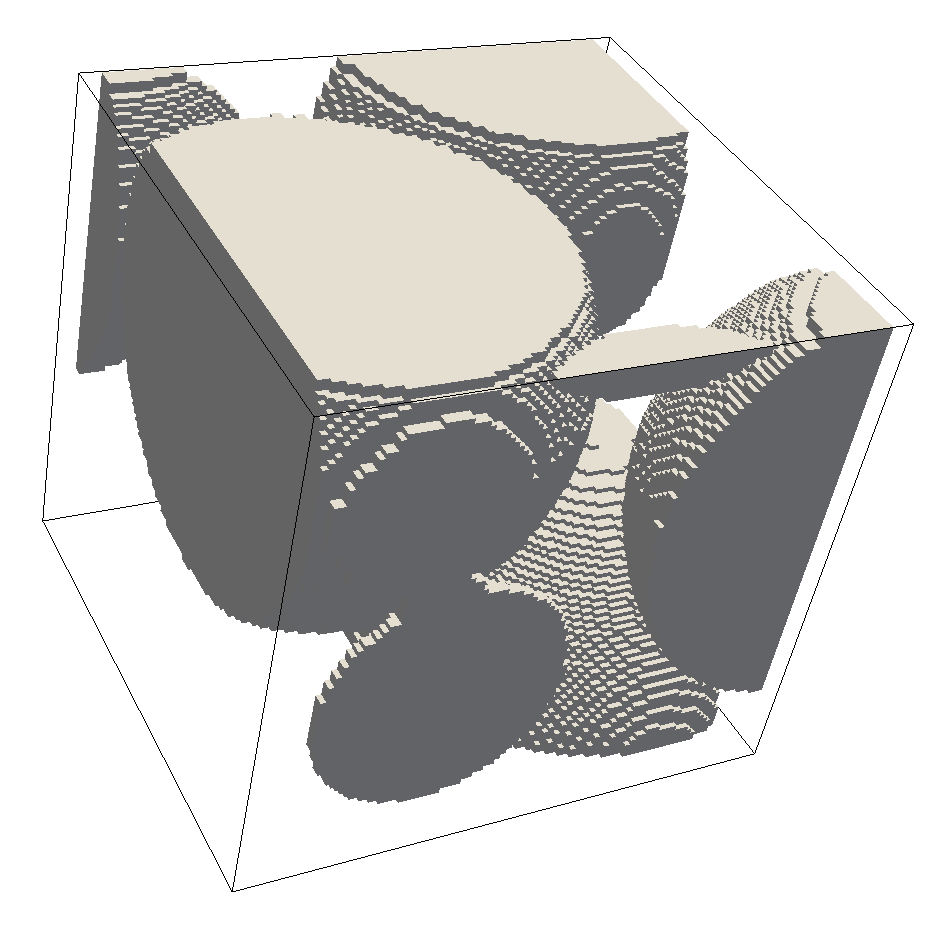}
		\caption{Irregular sphere packing}
		\label{fig:3d_omp_geo}
	\end{subfigure}%
	~   
	\begin{subfigure}[t]{0.45\textwidth}
		\centering
		\includegraphics[width=\textwidth]{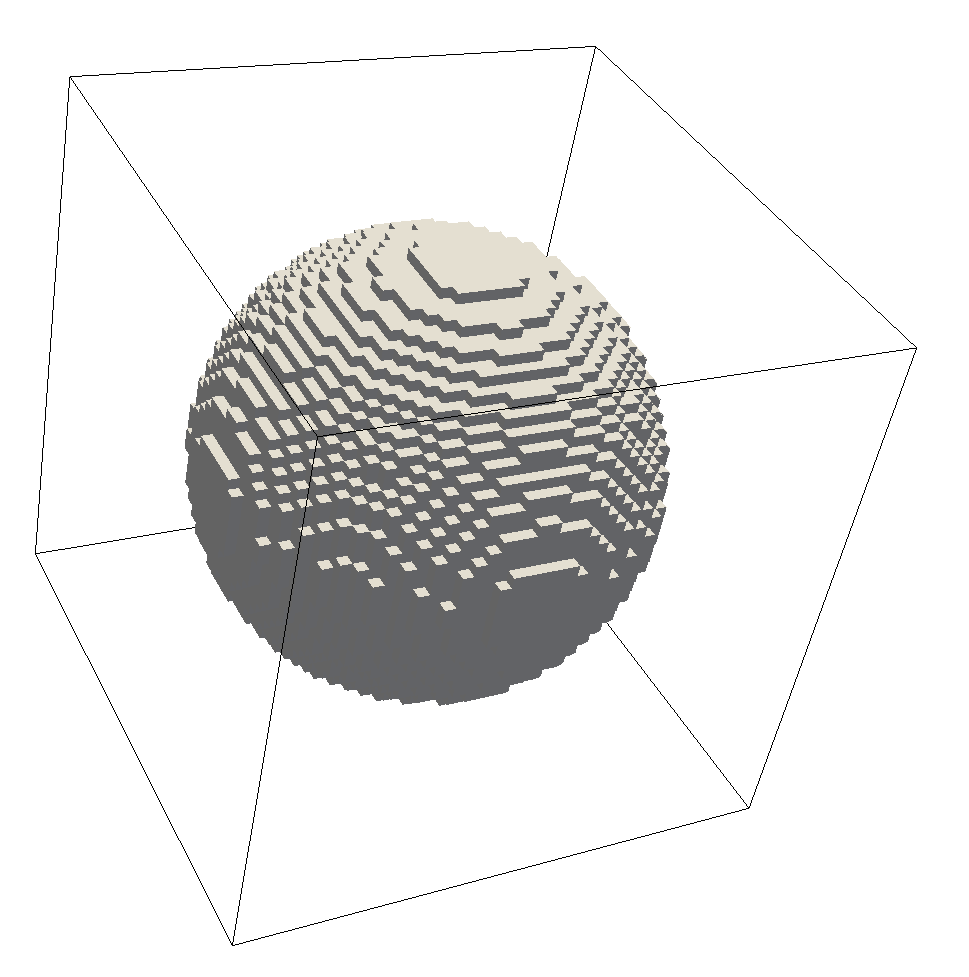}
		\caption{Cubic sphere packing}
		\label{fig:3d_weak_geo}
	\end{subfigure}%
	
	\caption{Relatively small computation domains for 3D models of porous media made by sphere (in grey color) packing. The spatial grid size $N_x$ and porosity $\epsilon$  are (a) $N_x=108\times100\times100$, $\epsilon=0.56$,  (b) $N_x=50\times50\times50$, $\epsilon=0.73$.}
	\label{fig:3d_geo_omp_weak}
\end{figure}

\section{Parallel performance}\label{sec:scaling performance}

Measuring the scaling performance of the OpenMP and MPI parallelisation individually allows us to better evaluate the overall efficiency of the multi-level parallel strategy. In this section, we demonstrate the scaling performance of each parallel level of the solver by excluding deterioration of convergence rate, i.e. comparing the wall clock time for an interval of $100$ iterations. We first present the results of fine-grained (the bottom-level OpenMP) scaling performance using only one MPI process, followed by the weak and strong scaling performance with multiple MPI processes (the top level of parallelization), each with multiple OpenMP threads. Finally, speedup of two-level MPI/OpenMP parallelization is compared with that of pure MPI parallelization.


\subsection{Fine-grained (the bottom-level OpenMP) scaling performance}
To evaluate the OpenMP speedup against the serial processing, the OpenMP threads are mapped to the cores of a single processor (the maximum number of threads is $12$) to avoid the performance drop caused by the non-uniform memory access (NUMA) across the two processors on the compute node.

\begin{figure*}[thbp]
	\centering
	\begin{subfigure}[t]{0.50\textwidth}
		\centering
		\includegraphics[width=\textwidth]{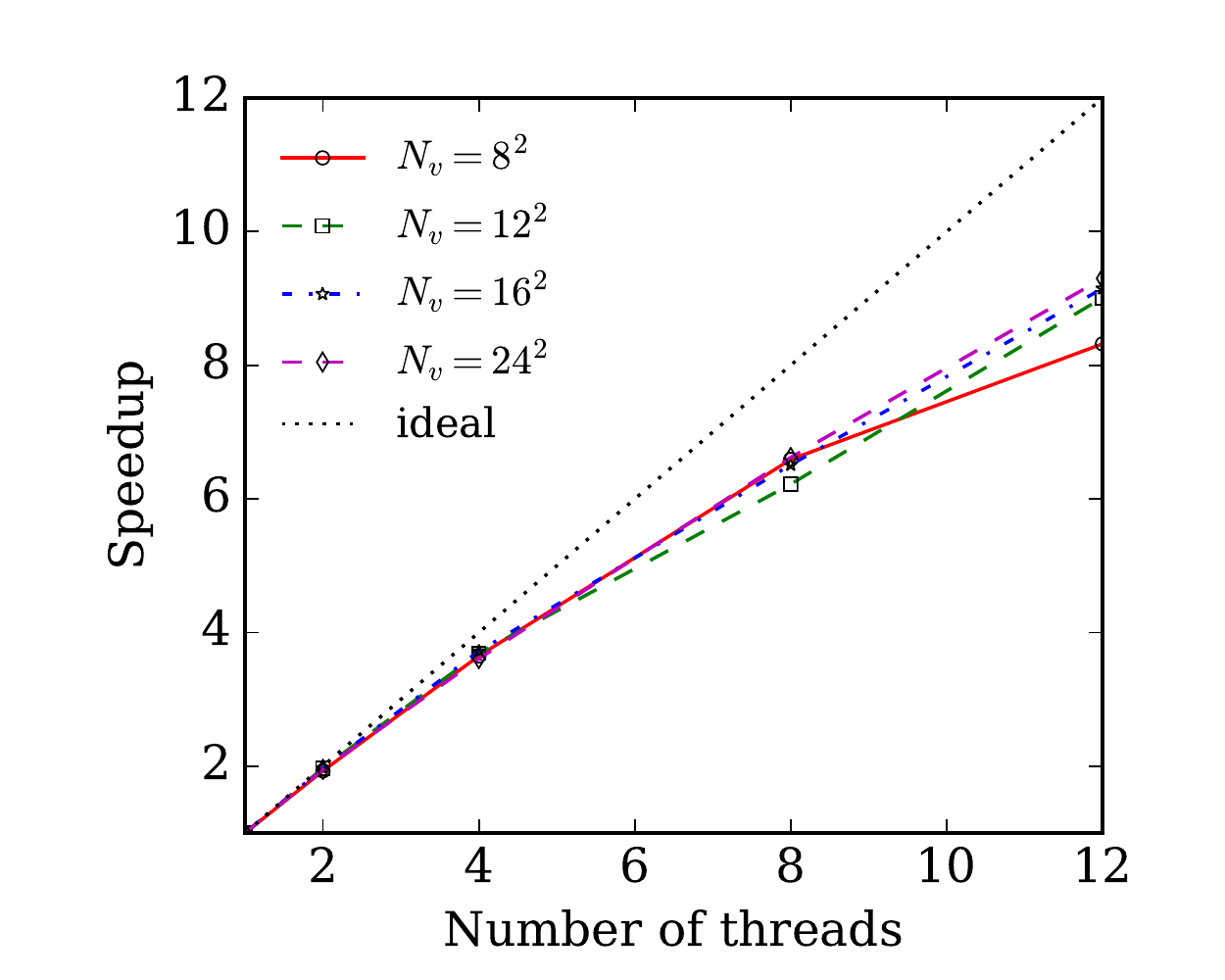}
		\caption{2D random cylinders}
		\label{fig:2d_omp_speedup}
	\end{subfigure}%
	~   
	\begin{subfigure}[t]{0.5\textwidth}
		\centering
		\includegraphics[width=\textwidth]{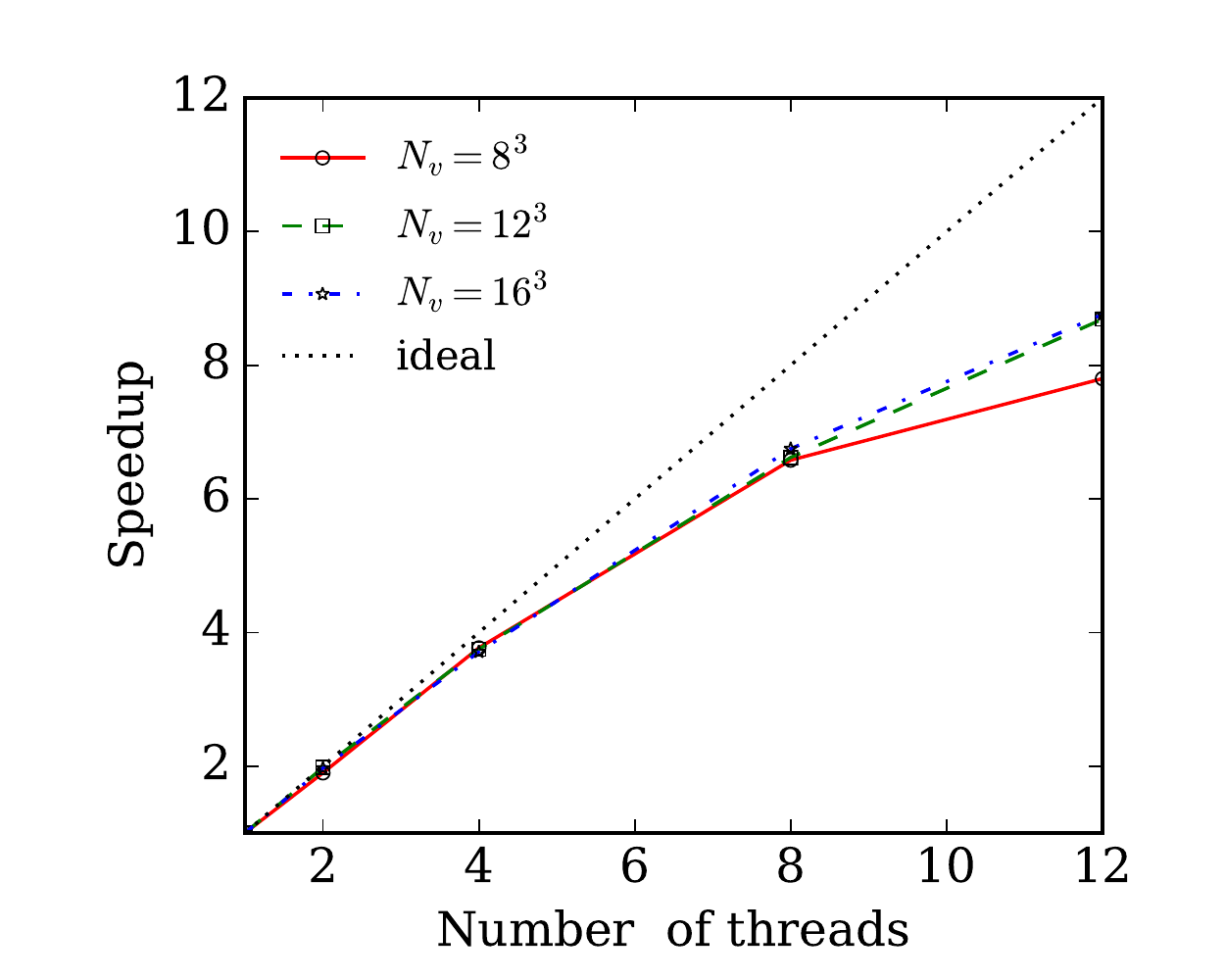}
		\caption{3D irregular sphere packing}
		\label{fig:3d_omp_speedup}
	\end{subfigure}%
	
	\caption{ Fine-grained (the bottom-level OpenMP) speedup for (a) the 2D porous model as shown in Fig.~\ref{fig:carpet_cylinders}(b), and (b) the 3D porous model as shown in Fig.~\ref{fig:3d_geo_omp_weak}(a). Various velocity grid refinement $N_v$ are considered.}
	\label{fig:omp_speedup}
\end{figure*}

The speedup of OpenMP parallelization for the 2D porous model as illustrated in Fig.~\ref{fig:carpet_cylinders}(b) is plotted in Fig.~\ref{fig:omp_speedup}(a). Almost perfect linear speedup is observed for all the examined velocity grids until 4 OpenMP threads. It can be seen that speedup slightly increases with the velocity grid refinement. In the case of $12$ OpenMP threads, the speedup varies from $8.32$ to $9.30$ with the velocity grids of $N_v=8^2$ and $N_v=24^2$, respectively.  

Figure~\ref{fig:omp_speedup}(b) plots the speedup of OpenMP parallelization for the 3D porous model as shown in Fig.~\ref{fig:3d_geo_omp_weak}(a), which is a portion extracted from irregular sphere packing as shown in Fig.~\ref{fig:singlesphere_beadpack}(b). Similar to the 2D case, speedup is almost linear until 4 OpenMP threads. In the case of $12$ OpenMP threads, the speedup increases slightly from $7.80$ to $8.76$ with the velocity grid refinement from $N_v=8^3$ to $N_v=16^3$. 

\subsection{Coarse-grained (the top-level MPI) scaling performance}
In order to measure the MPI scalability, the number of OpenMP threads for each MPI process is fixed at $12$, and each compute node is assigned to $2$ MPI processes, running on the two processors. All MPI scalabilities are assessed against $1$ compute node apart from the case of 3D strong scaling, where $4$ compute nodes are used as the baseline due to memory requirement.

\begin{figure*}[thbp]
	\centering
	\begin{subfigure}[t]{0.50\textwidth}
		\centering
		\includegraphics[width=\textwidth]{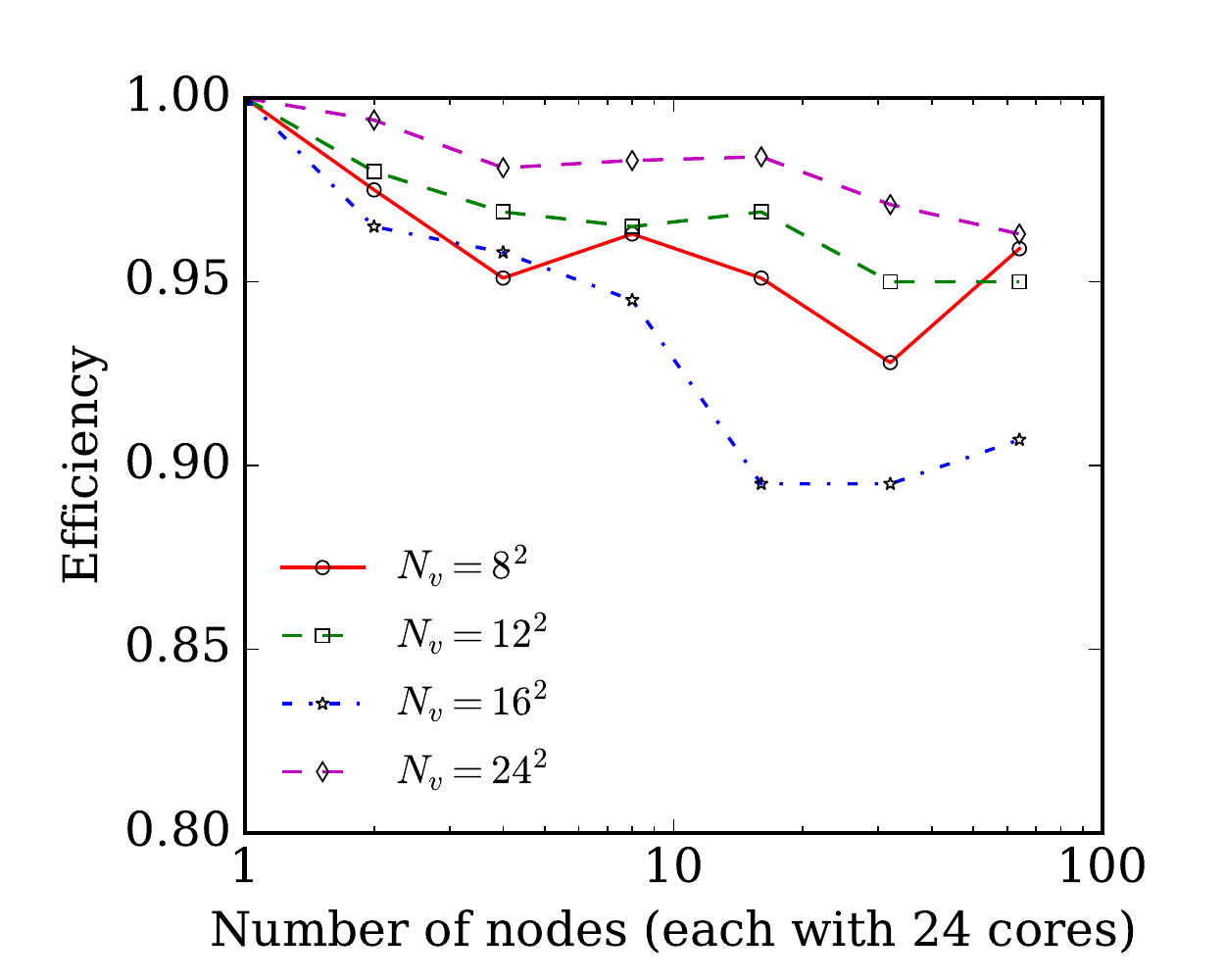}
		\caption{2D Sierpinski carpet}
		\label{fig:2d_weak_scaling}
	\end{subfigure}%
	~   
	\begin{subfigure}[t]{0.5\textwidth}
		\centering
		\includegraphics[width=\textwidth]{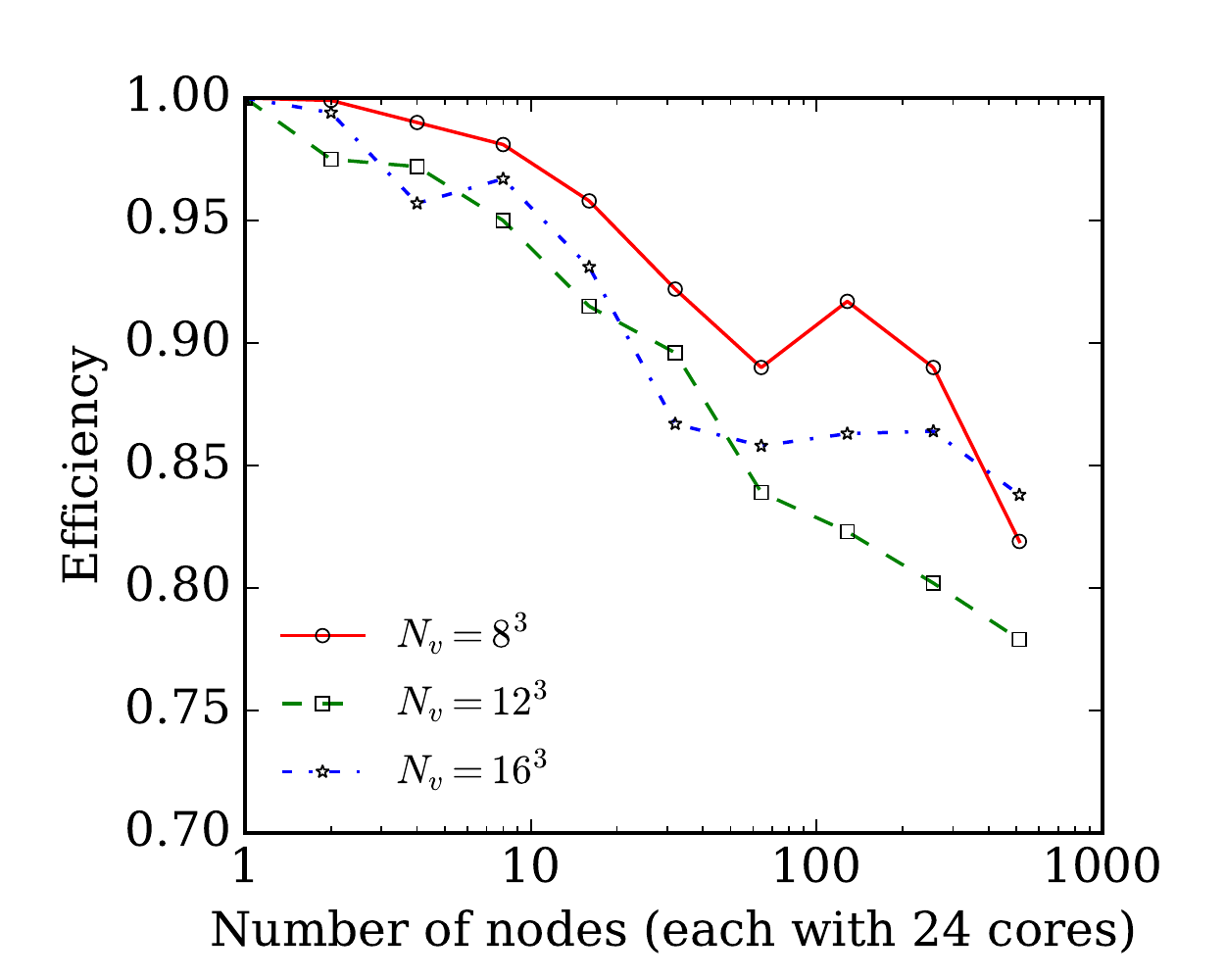}
		\caption{3D cubic sphere packing}
		\label{fig:3d_weak_scaling}
	\end{subfigure}%
	
	\caption{ The coarse-grained (the top-level MPI) efficiency of weak scaling, in which each MPI process handles a subdomain (a) as shown in  Fig.~\ref{fig:carpet_cylinders}(a) but with a coarser resolution $N_x=270\times270$; and (b) as shown in Fig.~\ref{fig:3d_geo_omp_weak}(b). Various velocity grid refinement $N_v$ are considered.}
	\label{fig:weak_scaling}
\end{figure*}

Weak scaling examines parallel performance by assuming that the spatial domain increases at the same rate as the number of cores, i.e. the workload per subdomain is equally fixed. Figure~\ref{fig:weak_scaling}(a) shows the weak scaling efficiency of the 2D solver, in which each MPI process manages a spatial subdomain as illustrated in Fig.~\ref{fig:carpet_cylinders}(a) with a coarse resolution of $N_x=270\times270$ pixels. The weak scaling efficiency only slightly decreases with increasing number of MPI processes. Very good parallel efficiency of about $94\%$ is observed with a total of $128$ MPI processes ($1536$ cores). In Fig.~\ref{fig:weak_scaling}(b), similar observation is found for the 3D solver, in which each MPI process handles a spatial subdomain as shown in Fig.~\ref{fig:3d_geo_omp_weak}(b). The weak scaling efficiency gradually declines to $86\%$ for the total of $128$ MPI processes ($1536$ cores) and $81\%$ for the $1024$ MPI processes ($12288$ cores). It is worth noting that speedup of $420$ is obtained with $512$ compute nodes for the grid size of $1.1\times10^{9}\cdot 4.1\times10^{3}$ as examined on this weak scaling. High efficiency of the weak scaling indicates the solvers' capability for a large number of voxels of homogeneous porous media.       

\begin{figure*}[thbp]
\centering
\begin{subfigure}[t]{0.50\textwidth}
\centering
\includegraphics[width=\textwidth]{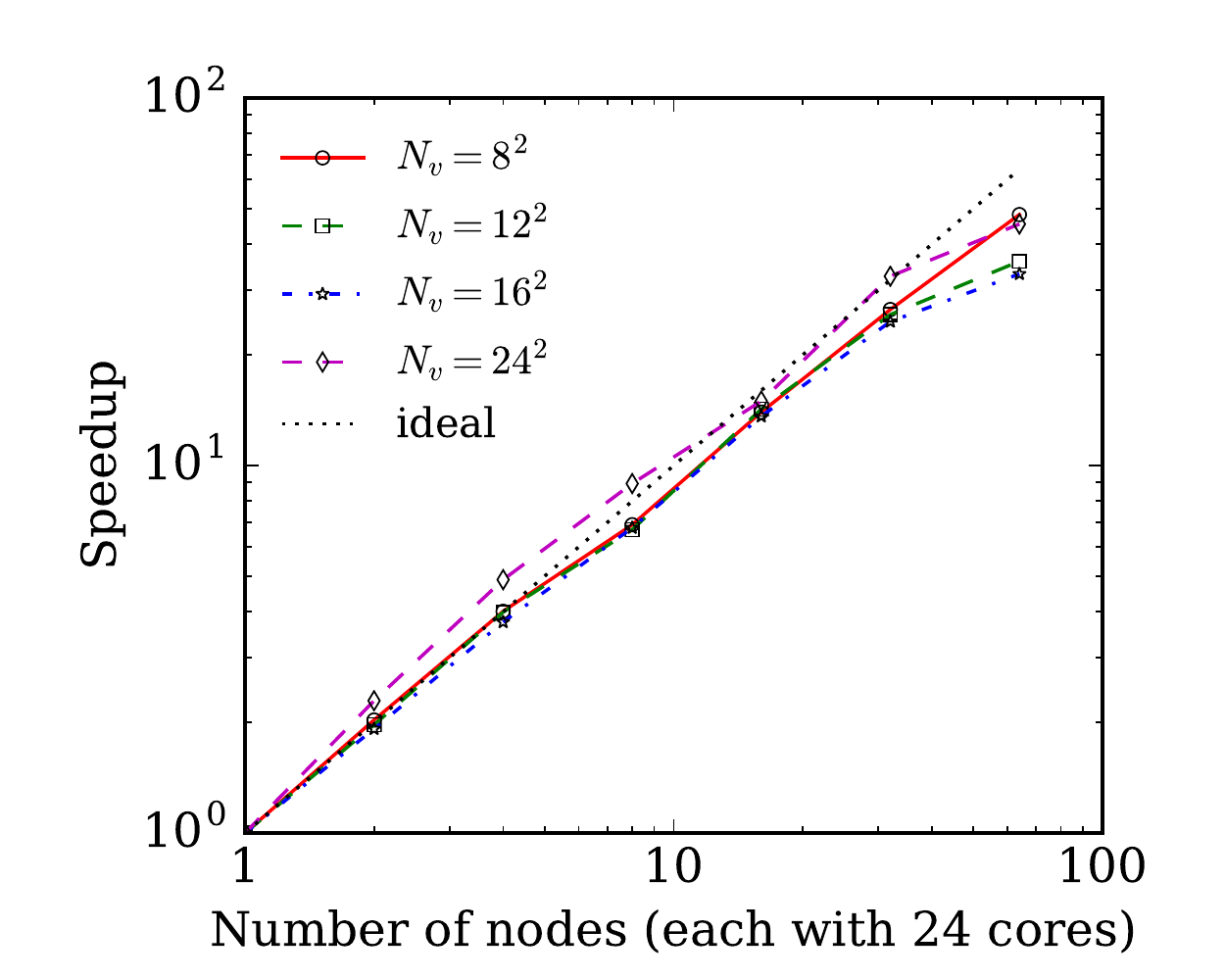}
\caption{2D random cylinders}
\label{fig:2d_strong_scaling}
\end{subfigure}%
~   
\begin{subfigure}[t]{0.5\textwidth}
\centering
\includegraphics[width=\textwidth]{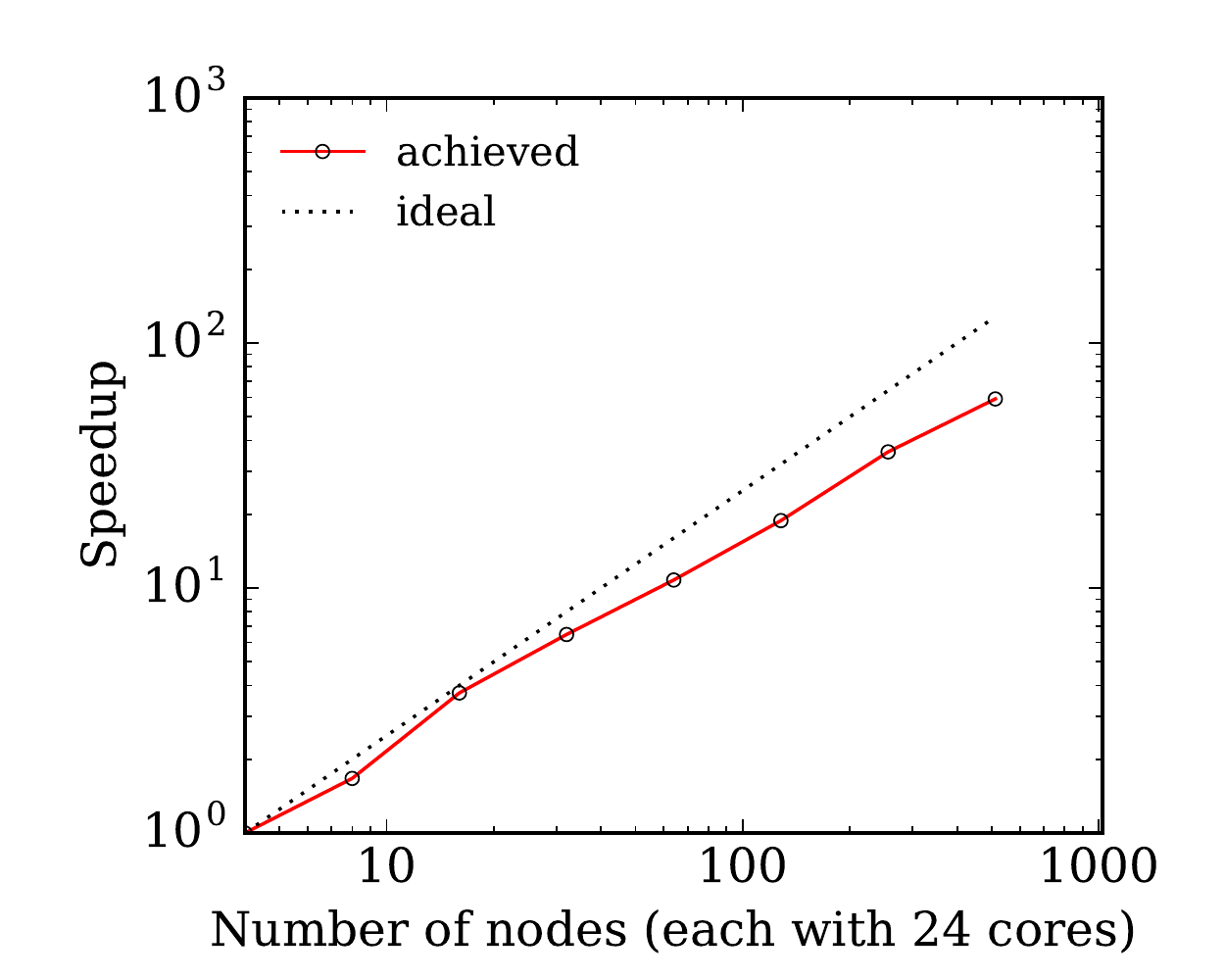}
\caption{3D irregular sphere packing}
\label{fig:3d_strong_scaling}
\end{subfigure}%

\caption{The coarse-grained (the top-level MPI) speedup of strong scaling for (a) the 2D porous model as shown in Fig.~\ref{fig:carpet_cylinders}(b); and (b) the 3D porous model as shown in Fig.~\ref{fig:singlesphere_beadpack}(b). Velocity grid of $N_v=8^2,12^2,16^2,24^2$ are considered in (a), and $N_v=8^3$ in (b).}
\label{fig:strong_scaling}
\end{figure*}

On the other hand, strong scaling examines the parallel performance when the computational domain remains unchanged as the number of cores increases. Figs.~\ref{fig:strong_scaling}(a) and \ref{fig:strong_scaling}(b) demonstrate the strong scaling speedup of the 2D and 3D solvers on the spatial domains illustrated in Figs.~\ref{fig:carpet_cylinders}(b) and \ref{fig:singlesphere_beadpack}(b), respectively. 
It can be seen that the 2D strong scaling speedup is close to ideal linear one until $64$ MPI processes ($768$ cores), where the efficiency is around $86\%$, and the efficiency falls to $64\%$ at $128$ MPI processes ($1536$ cores). In the 3D case, strong scaling speedup keeps increasing even beyond $12288$ cores, but the efficiency slumps from $81\%$ to $67\%$ as the number of MPI processes increases from $64$ to $128$ ($768$ to $1536$ cores).
The deterioration in strong scaling performances can be explained by load imbalance among the MPI processes, which is directly related to 
imhomogeneity of the porous models.
Consider the cases of $128$ MPI processes, where the spatial domains are decomposed into $16 \times 8$ rectangular subdomains for the 2D random cylinders as shown in Fig.~\ref{fig:carpet_cylinders}(b) or $4 \times 4 \times 8$ cuboid subdomains for the 3D irregular sphere packing as shown in Fig.~\ref{fig:singlesphere_beadpack}(b). The porosity $\epsilon$ of each subdomain varies from $0.313$ to $0.909$ for the 2D case ($\epsilon=0.60$ on average) or from $0.090$ to $0.706$ for the 3D case ($\epsilon=0.38$ on average), leading to strong load imbalance among the MPI processes. Improving load balance for heterogeneous porous media is certainly worth further studies, e.g. using the standard graph  partitioners such as METIS~\cite{Karypis:1998}, PT-SCOTCH~\cite{Chevalier:2008} or developing suitable algorithms. 
\begin{figure*}[thbp]
	\centering
	
	\begin{subfigure}[t]{0.50\textwidth}
        \centering
        \includegraphics[width=\textwidth]{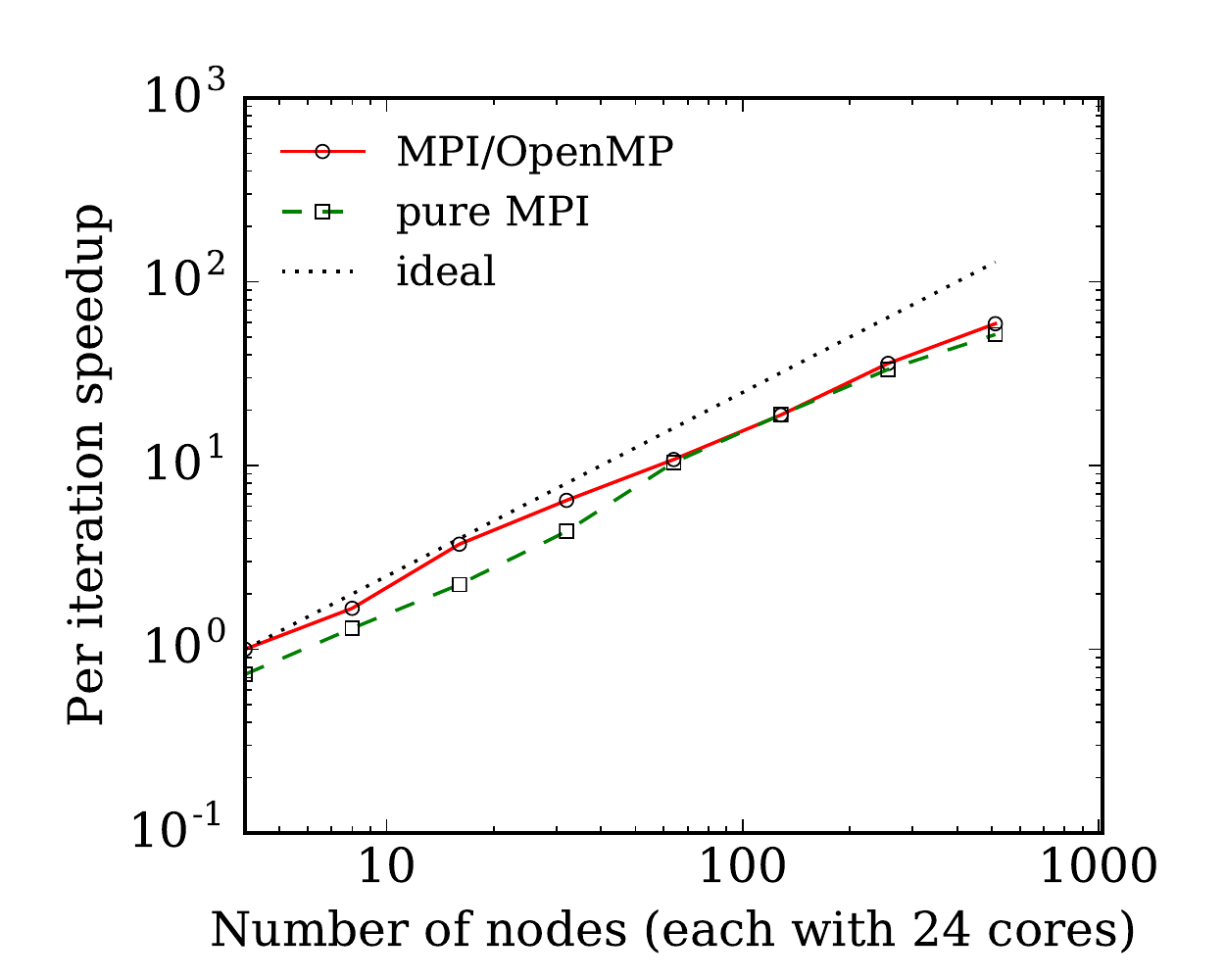}
        \caption{Measured on 100 iterations}
        \label{fig:purempi_hybrid_per_iteration}
    \end{subfigure}%
~   
	\begin{subfigure}[t]{0.50\textwidth}
        \centering
        \includegraphics[width=\textwidth]{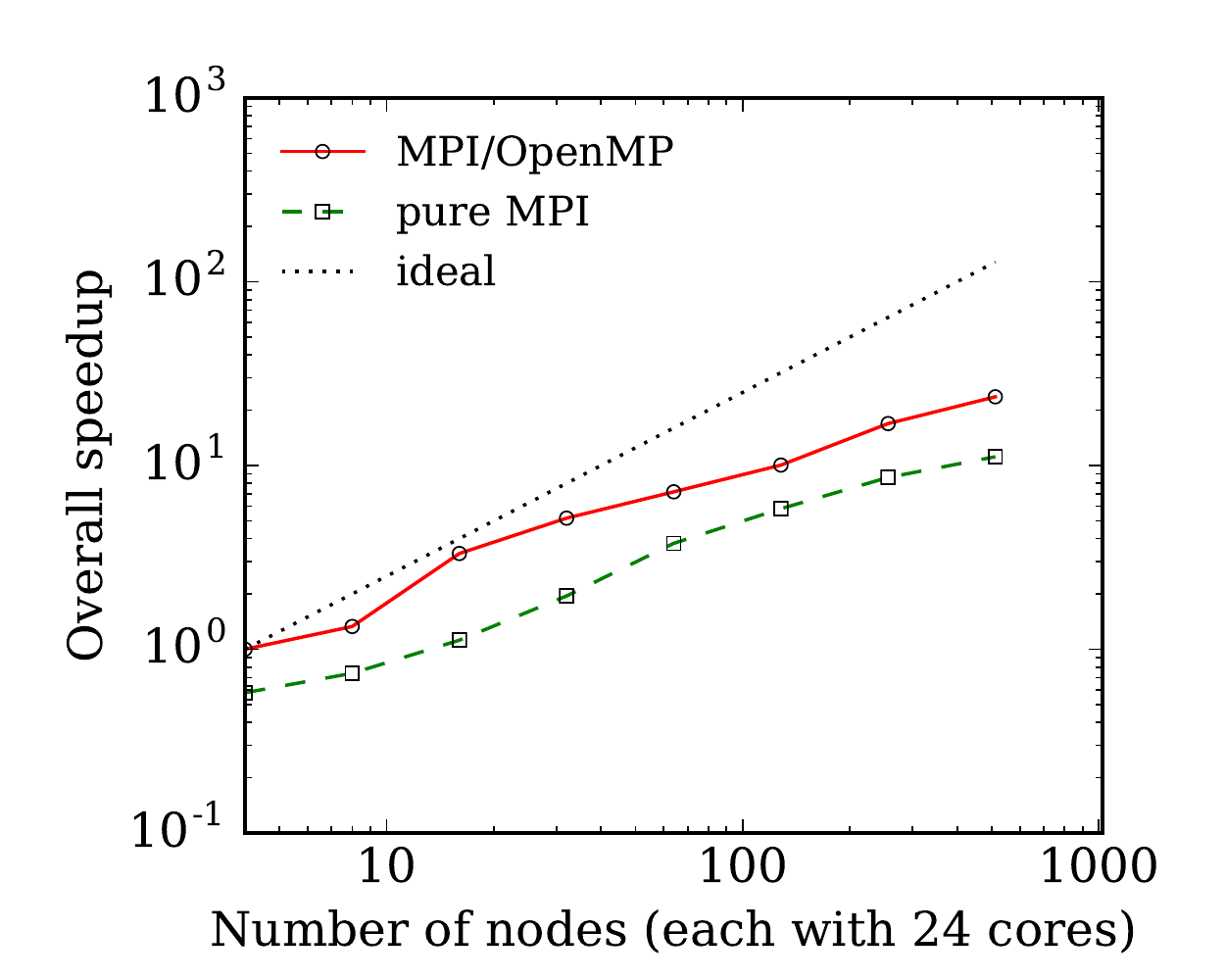}
        \caption{Measured on the whole simulation}
        \label{fig:purempi_hybrid_overall}
    \end{subfigure}%
	\caption{ Parallel speedup of strong scaling for the 3D irregular sphere packing as shown in Fig.~\ref{fig:singlesphere_beadpack}(b): multi-level MPI/OpenMP mode versus pure MPI mode. The velocity grid of $N_v=8^3$ is used for the case of $Kn=1$.}
	\label{fig:purempi_hybrid}
\end{figure*}

\subsection{Scaling performance of multi-level MPI/OpenMP versus pure MPI parallelism}
As mentioned in Sec.~\ref{sec:intro}, pure MPI parallelization commonly employs either velocity or spatial domain decomposition. But the velocity domain decomposition is not suitable for a relatively large spatial domain. It is interesting to compare the parallel performance of multi-level MPI/OpenMP approach adopted in this study with that of pure MPI approach using spatial domain decomposition. To enable pure MPI mode from the adopted multi-level parallel solvers, we disable OpenMP recognition by adding the compiler flag \textit{–hnoomp} in the Cray Programming Environment. 
The wall clock time of both $100$ iterations and whole simulations obtained by the multi-level and pure MPI modes for the case of 3D irregular sphere packing as shown in Fig.~\ref{fig:singlesphere_beadpack}(b) with $Kn=1$ is recorded.
The wall clock time of the multi-level mode on $4$ compute nodes is used as the baseline time for calculating speedup.
Figure~\ref{fig:purempi_hybrid}(a), in which the speedup on the interval of 100 iterations is reported, shows a relatively good scaling of both parallel modes until the maximum used resource, i.e. $12288$ cores. With the same resource, the multi-level mode is between $30\%$ and $60\%$ faster than the pure MPI mode until $768$ cores and becomes slightly faster beyond that number of cores. 
From Fig.~\ref{fig:purempi_hybrid}(b), when the deterioration of convergence rate is taken into account, it can be observed that even at $512$ compute nodes ($12288$ cores) the overall speedup of both the multi-level and pure MPI modes have not been saturated. However, with the same number of CPU cores, the overall speedup of multi-level mode is roughly $1.5$ to $3$ times better than that of the pure MPI mode. One main reason is that the multi-level mode has less number of spatial subdomains, i.e. MPI processes, than the pure MPI mode, resulting in faster convergence rate for the iterative scheme.    


\section{Conclusions and remarks}\label{sec:conclusion}


In this work, we have developed a high-performance DVM solver for steady 2D and 3D rarefied gas flows in porous media. 

While spatial domain decomposition is inevitable for practical large-scale pore-scale computations, the number of subdomains should be kept minimal to avoid deterioration of the convergence rate of the iterative scheme. Therefore, two-level MPI/OpenMP parallelization is proposed to improve parallel performance, where an additional parallel level (OpenMP) allows further speedup with the fixed number of subdomains or mitigates deterioration in convergence rate with the fixed CPU resource. The parallel scaling shows that the two-level parallel approach has significantly better performance than the commonly-used MPI approach for the same number of CPU cores. The deterioration of the convergence rate is found to become worse with increasing Knudsen number, but it can be mitigated by complex porous media.

The developed solver 
can enable 3D pore-scale simulations to predict the flow properties of porous media. Further investigations on optimization of grid partition is needed to improve scalability for heterogeneous porous media. 
This multi-level parallel approach, which is demonstrated with the BGK equation here, can be easily extended to solve other kinetic model equations.

\section*{Acknowledgements}
Financial support from the UK Engineering and Physical Sciences Research Council (EPSRC) under grant no.~EP/M021475/1 is gratefully acknowledged. L.~Zhu acknowledges the financial support from the Chinese Scholarship Council (CSC) during his visit to the UK (CSC Student no.~201606160050). L. Wu acknowledges the financial support from the RSE-NSFC joint project and Carnegie Research Incentive Grant. Computing time on the ARCHER is provided by the "UK Consortium on Mesoscale Engineering Sciences (UKCOMES)" under the UK EPSRC grant no.~EP/L00030X/1. 



\biboptions{square,comma,compress}
\bibliographystyle{elsarticle-num}
\bibliography{biblio}
\end{document}